\pdfoutput=1 
\documentclass[%
reprint,
 bibnotes,
 amsmath,
 amssymb,
 aps,
 prd,
 floatfix,
]{revtex4-1}
\usepackage[T1]{fontenc}
\usepackage[latin9]{inputenc}
\usepackage{float}
\usepackage{amsmath}
\usepackage{stackrel}
\usepackage{graphicx}
\usepackage{epstopdf}
\usepackage{setspace}
\usepackage{heppennames2}
\RequirePackage[colorlinks,citecolor=blue,urlcolor=blue,linkcolor=blue]{hyperref}
\usepackage{natbib}

\begin{document}

\title{Measurement of the Higgs boson mass and $\mathrm{e}^{+}\mathrm{e^{-}\rightarrow\mathrm{ZH}}$
cross section using $\mathrm{\mathrm{\mathrm{Z\rightarrow\mu^{+}\mu^{-}}}}$
and $\mathrm{Z\rightarrow\mathrm{e^{+}\mathrm{e^{-}}}}$ at the ILC}

\author{J.Yan,$^{1}$ S.Watanuki,$^{2}$ K.Fujii,$^{1}$ A.Ishikawa,$^{2}$
D. Jeans,$^{3}$ J. Strube,$^{4}$ J. Tian,$^{1}$ H.Yamamoto$^{2}$ }
\affiliation{
\textit{$^{1}$ High Energy Accelerator Research Organization (KEK), Tsukuba 305-0801, Japan}\\
\textit{$^{2}$ Department of Physics, Tohoku University, Sendai 980-8578, Japan}\\
\textit{$^{3}$ Department of Physics, Graduate School of Science, The University of Tokyo, 7-3-1 Hongo, Bunkyo-ku, Tokyo 113-0033, Japan}\\
\textit{$^{4}$ Pacific Northwest National Laboratory, 902 Battelle Boulevard P.O. Box 999, MSIN J4-60 Richland, WA 99352, U.S.A.}
}
\begin{abstract}
This paper presents a full simulation study of the measurement of
the production cross section ($\sigma_{\mathrm{ZH}}$) of the Higgsstrahlung
process $\mathrm{e^{+}e^{-}\rightarrow ZH}$ and the Higgs boson mass
($M_{\mathrm{H}}$) at the International Linear Collider (ILC), using
events in which a Higgs boson recoils against a Z boson decaying into
a pair of muons or electrons. The analysis is carried out for three
center-of-mass energies $\sqrt{s}$ = 250, 350, and 500 GeV, and two
beam polarizations $\mathrm{e_{L}^{-}e_{R}^{+}}$ and $\mathrm{e_{R}^{-}e_{L}^{+}}$,
for which the polarizations of $\mathrm{e^{-}}$ and $\mathrm{e^{+}}$
are $\left(P\mathrm{e^{-}},P\mathrm{e^{+}}\right)$ =($-$80\%, +30\%)
and (+80\%, $-$30\%), respectively. Assuming an integrated luminosity
of 250 $\mathrm{fb^{-1}}$ for each beam polarization at $\sqrt{s}$
= 250 GeV, where the best lepton momentum resolution is obtainable,
$\sigma_{\mathrm{ZH}}$ and $M_{\mathrm{H}}$ can be determined with
a precision of 2.5\% and 37 MeV for $\mathrm{e_{L}^{-}e_{R}^{+}}$
and 2.9\% and 41 MeV for $\mathrm{e_{R}^{-}e_{L}^{+}}$, respectively.
Regarding a 20 year ILC physics program, the expected precisions for
the $\mathrm{HZZ}$ coupling and $M_{\mathrm{H}}$ are estimated to
be 0.4\% and 14 MeV, respectively. The event selection is designed
to optimize the precisions of $\sigma_{\mathrm{ZH}}$ and $M_{\mathrm{H}}$
while minimizing the bias on the measured $\sigma_{\mathrm{ZH}}$
due to discrepancy in signal efficiencies among Higgs decay modes.
For the first time, model independence has been demonstrated to a
sub-percent level for the $\sigma_{\mathrm{ZH}}$ measurement at each
of the three center-of-mass energies. The results presented show the
impact of center-of-mass energy and beam polarization on the evaluated
precisons and serve as a benchmark for the planning of the ILC run
scenario.
\end{abstract}
\pacs{Valid PACS appear here}
\maketitle

\section{Introduction}

It is one of the most important missions of high energy particle physics
to uncover the physics behind electroweak symmetry breaking (EWSB).
The discovery of the Standard Model (SM)-like Higgs boson at the Large
Hadron Collider (LHC) in 2012 \cite{Aad:2012tfa,Chatrchyan:2012xdj} proved the basic idea
of the SM that the vacuum filled with the Higgs condensate broke the
electroweak symmetry. The SM assumes one doublet of complex scalar
fields for the Higgs sector. However, apart from the fact that it
is the simplest, there is no reason to prefer the Higgs sector in
the SM over any other model that is consistent with experiments. Moreover,
the SM does not explain why the Higgs field became condensed in vacuum.
To answer this question, we need physics beyond the SM (\textquotedblleft BSM\textquotedblright )
which necessarily alters the properties of the Higgs boson. Each new
physics model predicts its own size and pattern of the deviations
of Higgs boson properties from their SM predictions. In order to discriminate
these new physics models, we need to measure with high precision as
many types of couplings as possible and as model independently as
possible. Because the deviations predicted by most new physics models
are typically no larger than a few percent, the coupling measurements
must achieve a precision of 1\% or better for a statistically significant
measurement. This level of sensitivity is available only in the clean
experimental environment of lepton colliders.

The International Linear Collider (ILC)~\cite{Behnke:2013xla} is a proposed
$\mathrm{e^{+}e^{-}}$ collider covering center-of-mass energy range
of 200 to 500 GeV, with expandability to 1 TeV. Among the most important
aspects of its physics program~\cite{Fujii:2015jha} are the measurements
of Higgs couplings with unprecedented precision so as to find their
deviations from the SM and match their deviation pattern with predictions
of various new physics models.

Most of the Higgs boson measurements at the LHC are of cross section
times branching ratio (BR). This is also true at the ILC with one
important exception, the measurement of the absolute size of an inclusive
Higgs production cross section by applying the recoil technique to
the Higgsstrahlung process $\mathrm{e^{+}e^{-}\rightarrow ZH}$. The
recoil technique involves measuring only the momenta of the decay
products of the Z boson which recoils against the Higgs boson, and
hence in principle is independent of the Higgs decay mode. The measurement
of this cross section $\sigma_{\mathrm{ZH}}$ is indispensable for
extracting the branching ratios, the Higgs total width, and couplings
from cross section times branching ratio measurements. The recoil
technique, which is only possible at a lepton collider owing to the
well-known initial state, is applicable even if the Higgs boson decays
invisibly and hence allows us to determine $\sigma_{\mathrm{ZH}}$
in a completely model independent way.
The recoil technique also provides one of the most precise measurements
of the Higgs boson mass ($M_{\mathrm{H}}$), which is necessary for
estimating the phase space factor for the $\mathrm{HWW^{*}}$ decay
to extract the Higgs total width.

Especially high precision measurements of $\sigma_{\mathrm{ZH}}$
and $M_{\mathrm{H}}$ are possible by applying the recoil technique
to Higgsstrahlung events where the Z boson decays to a pair of electrons
or muons, which profits from excellent tracking momentum resolution
and relatively low background levels. Furthermore, in this channel
model independence for the measurement of $\sigma_{\mathrm{ZH}}$
can be demonstrated in practice.

This paper reports a study which evaluates the performance of measuring
$\sigma_{\mathrm{ZH}}$ and $M_{\mathrm{H}}$ using the Higgsstrahlung
process with a Z boson decaying into a pair of electrons or muons
$\mathrm{e^{+}e^{-}\rightarrow ZH\rightarrow\mathrm{l^{+}l^{-}}H}$
( $\mathrm{l}$ = e or $\mu$). One of the major purposes of this
study is to quantify the impact of center of mass energy and beam
polarization on the precision of $\sigma_{\mathrm{ZH}}$ and $M_{\mathrm{H}}$;
the analysis is carried out for three center-of-mass energies (250,
350, and 500 GeV), as well as two beam polarizations $\left(P\mathrm{e}^{-},P\mathrm{e}^{+}\right)$
=($-$80\%, +30\%) and (+80\%, $-$30\%), which will be denoted as
$\mathrm{e_{L}^{-}}\mathrm{e_{R}^{+}}$ and $\mathrm{e_{R}^{-}}\mathrm{e_{L}^{+}}$,
respectively.\footnote{A similar leptonic recoil analysis has previously been performed for
$\sqrt{s}$ = 250 GeV\cite{Abe:2010aa}.} Unless otherwise specified, the total integrated luminosity is assumed
as follows: For each beam polarization 250 $\mathrm{fb^{-1}}$, 333
$\mathrm{fb^{-1}}$, and 500 $\mathrm{fb^{-1}}$ are accumulated for
$\sqrt{s}$ = 250, 350, and 500 GeV, respectively. The H20 program
\cite{Barklow:2015tja}, one of the currently proposed ILC run scenarios which
covers startup, energy stages, and a luminosity upgrade, designates
that during a 20 year period, a total of 2000, 200, and 4000 $\mathrm{fb^{-1}}$
will be accumulated at $\sqrt{s}$= 250, 350, and 500 GeV, respectively.
The analysis results in this paper will be scaled to the luminosities
of the H20 program, and will impact the planning of future updates
of the run scenario.

The model-independence of the leptonic recoil technique has been evaluated in the context of previous
high-energy $\Pep\Pem$-colliders~\cite{GarciaAbia:1999kv}.
This paper demonstrates for the first time that the bias due to Higgs decay mode-dependence can be kept at the level
well below the expected statistical uncertainty in the H20 scenario
without sacrificing signal selection efficiency\footnote{\cite{Thomson:2015jda} presents an analysis using hadronic decays
of the Z boson at a center-of-mass energy of 350 GeV, as well as the
Higgs decay model independence of the $\sigma_{\mathrm{ZH}}$ obtained
through this analysis.}.

This paper is structured as follows: Section \ref{sec:Higgs-Recoil-Mass}
explains the recoil measurement; Section \ref{sec:Analysis} introduces
the simulation tools, the ILC detector concept, and the signal and
physics background processes; Section \ref{sec:Event-Selection} presents
the methods of data selection; Section \ref{sec:Extraction} gives
the methods for extracting $\sigma_{\mathrm{ZH}}$ and $M_{\mathrm{H}}$,
and discusses their expected precisions; Section \ref{sec:ModeBias}
demonstrates the model independence of the analysis; Section
\ref{sec:Summary-and-Conclusion} summarizes the analysis and concludes
the paper.

\section{Higgs Boson Measurements using the Recoil Technique}
\label{sec:Higgs-Recoil-Mass}

\begin{figure}
\centering
\includegraphics[width=.32\textwidth]{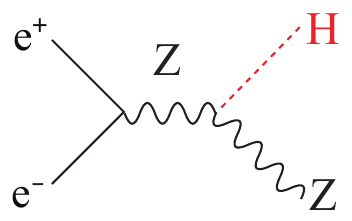}\hfill
\includegraphics[width=.32\textwidth]{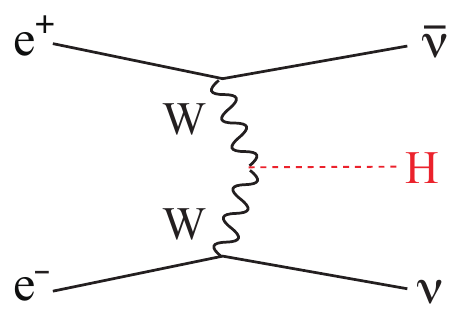}\hfill
\includegraphics[width=.32\textwidth]{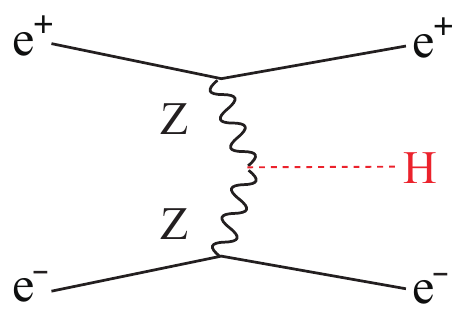}
\caption{The lowest order Feynman diagrams of the three major Higgs production
processes at the ILC: (top) Higgsstrahlung process $\mathrm{e^{+}e^{-}\rightarrow ZH}$,
(center) WW fusion process $\mathrm{e^{+}e^{-}\rightarrow\nu\overline{\nu}H}$,
and (bottom) ZZ fusion process $\mathrm{e^{+}e^{-}\rightarrow e^{+}e^{-}H}$.
}
\label{Feynmann}
\end{figure}

\begin{figure}
\centering
\includegraphics[width=.5\textwidth]{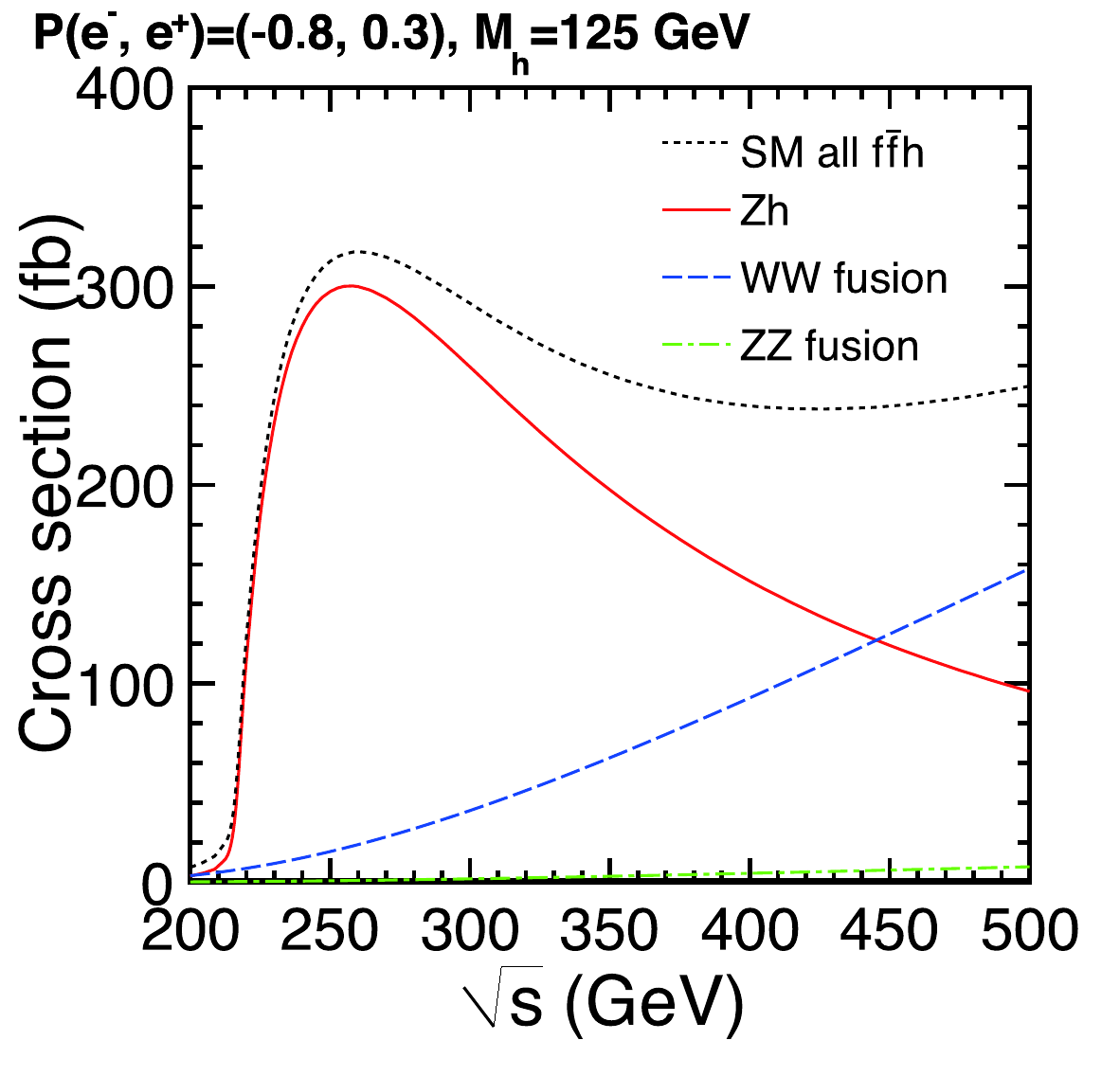}

\caption{The Higgs production cross section as a function of $\sqrt{s}$ assuming
$M_{\mathrm{H}}$=125 GeV for the following Higgs production processes:
Higgsstrahlung (solid), WW fusion (dashed), and ZZ fusion (dotted).
(Figure taken from \cite{Behnke:2013xla}.) }
\label{crossSec}
\end{figure}

The major Higgs production processes at the ILC are Higgsstrahlung
and WW fusion, whose lowest order Feynman diagrams are illustrated
in Figure \ref{Feynmann}, along with the ZZ fusion process which
has a significantly smaller cross section than the other two processes
at ILC center-of-mass energies. Figure \ref{crossSec} shows the production
cross sections as a function of $\sqrt{s}$, assuming a Higgs boson
mass of 125 GeV. The Higgsstrahlung cross section peaks around $\sqrt{s}$
= 250 GeV, and decreases gradually as $\sim 1/s$, whereas the WW fusion
cross section increases with energy, exceeding the Higgsstrahlung
process at around 450 GeV.

The Higgsstrahlung process with a Z boson decaying into a pair of
electrons or muons: $\mathrm{e^{+}e^{-}\rightarrow ZH\rightarrow\mathrm{l^{+}l^{-}}H}$
( $\mathrm{l}$ = e or $\mu$) will be hereafter referred to as $\mathrm{e^{+}e^{-}H}$
and $\mathrm{\mu^{+}\mu^{-}H}$, respectively. The leptonic
recoil technique is based on the Z boson identification by the invariant
mass of the dilepton system being consistent with the Z boson mass,
and the reconstruction of the mass of the rest of the final-state
system recoiling against the Z boson ($M_{\mathrm{rec}}$), corresponding
to the Higgs boson mass, which is calculated as

\begin{equation}
M_{\mathrm{rec}}^{2}=\left(\sqrt{s}-E_{\mathrm{l^{+}l^{-}}}\right)^{2}-\left|\overrightarrow{p}_{\mathrm{l^{+}l^{-}}}\right|^{2}\:,\label{eq:kinematics1}
\end{equation}
where $E_{\mathrm{l^{+}l^{-}}}\equiv E_{\mathrm{l^{+}}}+E_{\mathrm{l^{-}}}$
and $\overrightarrow{p}_{\mathrm{l^{+}l^{-}}}\equiv\overrightarrow{p}_{\mathrm{l}^{+}}+\overrightarrow{p}_{\mathrm{l}^{-}}$
are the energy and momentum of the lepton pair from Z boson decay.
The $M_{\mathrm{rec}}$ calculated using Equation \ref{eq:kinematics1}
is expected to form a peak corresponding to Higgs boson production.
From the location of the $M_{\mathrm{rec}}$ peak and the area beneath
it the Higgs boson mass and the signal yield can be extracted. The
signal selection efficiency, and hence the production cross section
is, in principle, independent of how the Higgs boson decays, since
only the leptons from the Z decay need to be measured in the recoil
technique. In practice, however, this is not guaranteed since there
is a possibility of confusion between the leptons from the Z boson
decay and those from the Higgs boson decay. It is thus an important
part of this study to demonstrate an analysis in which the signal
efficiency is indeed independent of assumptions regarding Higgs boson
decay.

\section{Analysis Framework, Detector Simulation, and Event Generation}
\label{sec:Analysis}

\subsection{Analysis Framework}

This study used the simulation and reconstruction tools contained
in the software package ILCSoft v01-16 \cite{ilcsoft:webpage}. All parameters
of the incoming beams are simulated with the GUINEA-PIG package \cite{Schulte:331845,Schulte:382453}
and the beam spectrum, including beamstrahlung and initial state radiation (ISR), are explicitly
taken into consideration based on the parameters in the TDR. The beam
crossing angle of 14 mrad in the current ILC design is taken into
account. The $\mathrm{\mu^{+}\mu^{-}H}$, $\mathrm{e^{+}e^{-}H}$,
and SM background Monte Carlo (MC) samples (see Section \ref{sub:Signal-and-Background}
for details) are generated using the WHIZARD 1.95 \cite{Kilian:2007gr}
event generator. The input mass of the Higgs boson is 125 GeV, and
its SM decay branching ratios are assumed \cite{Heinemeyer:2013tqa}.
The model for the parton shower and hadronization is taken from PYTHIA
6.4 \cite{Sjostrand:2001yu}. The generated events are passed through the ILD
\cite{Behnke:2013lya} simulation performed with the MOKKA\cite{MoradeFreitas:2002kj} software
package based on GEANT4\cite{Agostinelli:2002hh}. Event reconstruction is performed
using the Marlin\cite{Gaede:2006pj} framework. The PandoraPFA\cite{2009NIMPA.611...25T}
algorithm is used for calorimeter clustering and the analysis of track
and calorimeter information based on the particle flow approach.

\subsection{The ILD Concept}
\label{sub:The-ILD-Detector}

The International Large Detector (ILD) concept is one
of the two detectors being designed for the ILC. It features a hybrid
tracking system with excellent momentum resolution. The jet energy
resolution is expected to be better than 3\% for jets with energies
$\geq$ 100 GeV, thanks to its highly granular calorimeters optimized
for Particle Flow reconstruction. This section describes
the ILD sub-detectors important for this study.

The vertex detector (VTX), consisting of three double layers of extremely
fine Si pixel sensors with the innermost radius at 15 mm, measures
particle tracks with a typical spatial resolution of 2.8 $\mathrm{\mu m}$.
The hybrid tracking system consists of a time projection chamber (TPC)
which provides up to 224 points per track, excellent spatial resolution
of better than 100 $\mathrm{\mu m}$, and $dE/dx$ - based particle
identification, as well as Si-strip sensors placed in the barrel region
both inside and outside the TPC and in the end cap region outside the
TPC in order to further improve track momentum resolution. The tracking
system measures charged particle momenta to a precision of $\frac{\delta p_{t}}{p_{t}^{2}}=2\times10^{-5}$$\mathrm{\;GeV^{-1}}$.
Outside of the tracking system sits the ECAL, a Si-W sampling electromagnetic
calorimeter with an inner radius of 1.8 m, finely segmented $5\times5$
$\mathrm{mm^{2}}$ transverse cell size and 30 longitudinal layers
equivalent to 24 radiation lengths. The HCAL, a steel-scintillator
type hadronic calorimeter which surrounds the ECAL, has an outer radius
of 3.4 m, $3\times3$ $\mathrm{cm^{2}}$ transverse tiles, and 48
longitudinal layers corresponding to 5.9 interaction lengths. Radiation
hard calorimeters for monitoring the luminosity and quality of the
colliding beams are installed in the forward region. The tracking
system and calorimeters are placed inside a superconducting solenoid
which provides a magnetic field of 3.5 T. An iron yoke outside the
solenoid coil returns the magnetic flux, and is instrumented with
scintillator-based muon detectors.

\subsection{Signal and Background Processes}
\label{sub:Signal-and-Background}

\begin{figure}
\centering
\includegraphics[width=.32\textwidth]{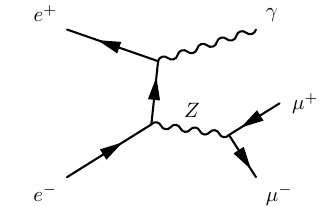} \hfill
\includegraphics[width=.32\textwidth]{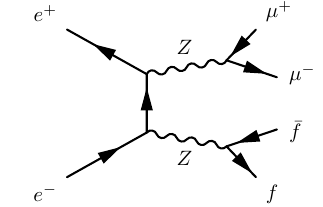} \hfill
\includegraphics[width=.32\textwidth]{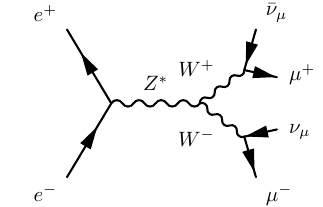}
\caption{The Feynman diagrams contributing to the major background processes
for the Higgs recoil analysis in the $\mathrm{\mu^{+}\mu^{-}H}$
channel: 2f\_l background with $\mu\mu$ in the final state and an
ISR photon (top), 4f\_sl background with ZZ as intermediate state
(center), 4f\_l background with WW as intermediate state (bottom).}
\label{fig:The-FeynmanBG}
\end{figure}

The Higgsstrahlung signal is selected by identifying a pair of prompt,
isolated, and oppositely charged muons or electrons with well-measurable
momentum whose invariant mass $M_{\mathrm{l^{+}l^{-}}}$ ($\mathrm{l}$=$\mathrm{e}$
or $\mu$) is close to the Z boson mass ($M_{\mathrm{Z}}$). The $\mathrm{\mu^{+}\mu^{-}H}$
and $\mathrm{e^{+}e^{-}H}$ channels are analyzed independently and
then statistically combined. Figure \ref{fig:The-FeynmanBG} shows
the Feynman diagrams of the dominant 4-fermion and 2-fermion processes.
Table \ref{tab:Processes-and-cross} gives the cross sections of signal
and major background processes assuming $M_{\mathrm{H}}$=125 GeV.
For each process, all SM diagrams are included at tree level.
These processes are grouped as follows from the perspective of finding
leptons in the final state:
\begin{itemize}
\item $\mathrm{l^{+}l^{-}}\mathrm{H}$ ($\mathrm{l}$=$\mathrm{e}$ or $\mu$)
: The Higgsstrahlung signal process with Z decaying to $\mathrm{l^{+}l^{-}}$.
The $\mathrm{e^{+}e^{-}H}$ channel contains an admixture of the ZZ
fusion process, which is removed at the early stages of the analysis.
\item 2-fermion leptonic (2f\_l): final states consisting of a charged lepton
pair or a neutrino pair. The intermediate states are Z or $\gamma^{*}$.
\item 4-fermion leptonic (4f\_l): final states of 4 leptons consisting of
mainly processes through ZZ and WW intermediate states. Those events
containing a pair of electrons or muons are a background of the $\mathrm{\mu^{+}\mu^{-}H}$
and $\mathrm{e^{+}e^{-}H}$ channels, respectively.
\item 4-fermion semi-leptonic (4f\_sl): final states of a pair of charged
leptons and a pair of quarks, consisting of mainly processes through
ZZ and WW intermediate states. In the former case, one Z boson decays
to a pair of charged leptons or neutrinos, and the other to quarks.
In the latter case, one W boson decays to a charged lepton and a neutrino
of the same flavor and the other to quarks.
\item 4(2)-fermion hadronic (4(2)f\_h): final states of 4 (2) quarks. Since
the probability of finding isolated leptons is very small for these
final states, these events are removed almost completely at the lepton
identification stage (see Section \ref{sub:Selection-of-Best}).
\end{itemize}
The analysis in this paper is conducted for the center-of-mass energies
250, 350, and 500 GeV, and two beam polarization $\mathrm{e_{L}^{-}e_{R}^{+}}$
and $\mathrm{e_{R}^{-}e_{L}^{+}}$. From Table \ref{tab:Processes-and-cross},
it can be seen that the signal cross sections for $\mathrm{e_{R}^{-}e_{L}^{+}}$
are smaller by a factor of 1.5 with respect to $\mathrm{e_{L}^{-}e_{R}^{+}}$.
The methods and performance of signal selection and background rejection
are presented in Section \ref{sec:Event-Selection}.

The Monte Carlo (MC) samples are generated for the cases in which
the polarizations of $\mathrm{e^{-}}$ and $\mathrm{e^{+}}$ are $\left(P\mathrm{e^{-}},P\mathrm{e^{+}}\right)$
=($-$100\%, +100\%) and (+100\%, $-$100\%). The standard samples
used in this paper are generated for signal and background processes
with the statistics as shown in Table \ref{tab:Processes-and-cross}.
Another type of signal sample is generated with high statistics of
more than 40k for each major SM Higgs decay mode, mainly for the purpose
of the model independence study in Section \ref{sec:ModeBias}. Unless
otherwise stated, the distributions shown in the following sections
are made using the standard samples and normalized to the assumed
integrated luminosities, cross sections, and polarizations.

\begin{table}
\centering
\caption{Cross sections and number of generated MC events ($N_{\mathrm{Gen}}$)
of signal and major background processes at each center-of-mass energy.
Cross sections, as calculated by the WHIZARD generator, are given for ILC beam polarizations
$\left(P\mathrm{e^{-}},P\mathrm{e^{+}}\right) = (\pm 80 \%, \mp 30\%)$,
while numbers of generated events are given for 100\% beam polarization.
\label{tab:Processes-and-cross}}

\begin{tabular}{|c|c|c|c|c|}
\hline
 & \multicolumn{2}{c|}{cross section} & \multicolumn{2}{c|}{$N_{\mathrm{Gen}}$}  \tabularnewline
\hline
$P\mathrm{e^{-}}$  & $-80\%$ & $+80\%$ & $-100\%$ & $+100\%$ \tabularnewline
$P\mathrm{e^{+}}$  & $+30\%$ & $-30\%$ & $+100\%$ & $-100\%$ \tabularnewline
\hline
\hline
 \multicolumn{5}{|c|}{$\sqrt{s}$ = 250 GeV}  \tabularnewline
\hline
$\mathrm{\mu^{+}\mu^{-}H}$  & 10.4 fb & 7.03 fb & 17.1k & 11.0k\tabularnewline
\hline
$\mathrm{e^{+}e^{-}H}$ & 10.9 fb & 7.38 fb & 17.6k & 11.2k\tabularnewline
\hline
2f\_l & 38.2 pb & 35.0 pb & 2.63M & 2.13M\tabularnewline
\hline
2f\_h & 78.1 pb & 46.2 pb & 1.75M & 1.43M\tabularnewline
\hline
4f\_l & 5.66 pb & 1.47 pb & 2.25M & 0.35M\tabularnewline
\hline
4f\_sl & 18.4 pb & 2.06 pb & 4.43M & 0.36M\tabularnewline
\hline
4f\_h & 16.8 pb & 1.57 pb & 2.50M & 0.24M\tabularnewline
\hline
total background & 157.1 pb & 86.3 pb & 13.6M & 4.51M\tabularnewline
\hline
\hline
 \multicolumn{5}{|c|}{$\sqrt{s}$ = 350 GeV}  \tabularnewline
\hline
$\mathrm{\mu^{+}\mu^{-}H}$  & 6.87 fb & 4.63 fb & 11.3k & 8.0k\tabularnewline
\hline
$\mathrm{e^{+}e^{-}H}$ & 10.24 fb & 6.68 fb & 17.9k & 9.0k\tabularnewline
\hline
2f\_l & 33.5 pb & 31.5 pb & 2.71M & 1.94M\tabularnewline
\hline
2f\_h & 38.6 pb & 23.0 pb & 1.60M & 0.89M\tabularnewline
\hline
4f\_l & 4.90 pb & 1.48 pb & 3.07M & 0.48M\tabularnewline
\hline
4f\_sl & 14.5 pb & 1.70 pb & 4.77M & 0.37M\tabularnewline
\hline
4f\_h & 12.6 pb & 1.11 pb & 2.49M & 0.22M\tabularnewline
\hline
total background & 104.1 pb & 58.7 pb & 14.6M & 3.89M\tabularnewline
\hline
%
%
\hline
 \multicolumn{5}{|c|}{$\sqrt{s}$ = 500 GeV}  \tabularnewline
\hline
$\mathrm{\mu^{+}\mu^{-}H}$  & 3.45 fb & 2.33 fb & 6.0k & 4.0k\tabularnewline
\hline
$\mathrm{e^{+}e^{-}H}$ & 11.3 fb & 7.11 fb & 15.0k & 7.5k\tabularnewline
\hline
2f\_l & 6.77 pb & 5.96 pb & 0.42M & 0.36M\tabularnewline
\hline
2f\_h & 19.6 pb & 11.7 pb & 1.51M & 0.84M\tabularnewline
\hline
4f\_l & 10.6 pb & 7.48 pb & 0.60M & 0.34M\tabularnewline
\hline
4f\_sl & 13.2 pb & 2.94 pb & 0.97M & 99.9k\tabularnewline
\hline
4f\_h & 8.65 pb & 0.74 pb & 0.69M & 18.0k\tabularnewline
\hline
total background & 58.9 pb & 28.8 pb & 4.18M & 1.65M\tabularnewline
\hline
\end{tabular}
\end{table}

\section{Analysis}
\label{sec:Event-Selection}

First, the signal events are selected by identifying a pair of leptons
($\mathrm{\mathrm{e^{+}\mathrm{e^{-}}}}$ or $\mu^{+}\mu^{-}$) produced
in the decay of the Z boson against which the Higgs recoils. Then
the recovery of final state radiation (FSR)/bremsstrahlung photons are performed. Finally
background events are rejected through a series of cuts on several
kinematic variables.

\subsection{Selection of Best Lepton Pair}
\label{sub:Selection-of-Best}

\subsubsection{Isolated Lepton Finder}
\label{sub:Isolated-Lepton-Finder}

Table \ref{LepID} summarizes the criteria for selecting an isolated
lepton. Here, $p_{\mathrm{track}}$ is the measured track momentum,
$E_{\mathrm{ECAL}}$ is the energy deposit in the ECAL, $E_{\mathrm{CAL,tot}}$
is the energy deposit in both ECAL and HCAL, $E_{\mathrm{yoke}}$
is the energy deposit inside the muon detector, and $d_{0}$ and $z_{0}$
are the transverse and longitudinal impact parameters. These criteria
are described as follows:
\begin{enumerate}
\item An electron deposits nearly all its energy in the ECAL while a muon
passes the ECAL and HCAL as a minimal ionizing particle. Therefore
$E_{\mathrm{ECAL}}$, $E_{\mathrm{CAL,tot}}$, and $p_{\mathrm{track}}$
are compared for each final state particle.
\item The leptons from $\tau$ decay or b/c quark jets are suppressed by
requirements on $d_{0}$ and $z_{0}$ with respect to their measurement
uncertainties.
\item In order to avoid selecting leptons in hadronic jets, the leptons
are required to have sufficient $p_{\mathrm{track}}$, and to satisfy
an isolation requirement based on a multi-variate double cone method \cite{JunpingTian2015:doubleCone}.
\end{enumerate}

\begin{table}
\centering
\caption{The criteria for the identification of isolated leptons ($\mu$ and
$e$). \label{LepID}}

\begin{tabular}{|c|c|}
\hline
 $\mu$ ID & $\mathrm{e}$ ID\tabularnewline
\hline
\hline
$p_{\mathrm{track}}>5\quad\mathrm{GeV}$ & $p_{\mathrm{track}}>5\quad\mathrm{GeV}$\tabularnewline
\hline
 $E_{\mathrm{CAL,tot}}/p_{\mathrm{track}}<0.3$ & $0.5<E_{\mathrm{CAL,tot}}/p_{\mathrm{track}}<1.3$\tabularnewline
\hline
 $E_{\mathrm{yoke}}<1.2\quad\mathrm{GeV}$ & $E_{\mathrm{ECAL}}/E_{\mathrm{CAL,tot}}>0.9$\tabularnewline
\hline
$\left|d_{0}/\delta d_{0}\right|<5$ & $\left|d_{0}/\delta d_{0}\right|<50$\tabularnewline
\hline
$\left|z_{0}/\delta z_{0}\right|<5$ & $\left|z_{0}/\delta z_{0}\right|<5$\tabularnewline
\hline
\end{tabular}
\end{table}

\subsubsection{Selection of the Best Lepton Pair}
\label{sub:pair}

For each event, two isolated leptons of the same flavor and opposite
charges are selected as the candidate pair for analysis. In this stage,
it is essential to distinguish a pair of leptons produced in the decay
of the Z boson recoiling against the Higgs boson (``correct pair'')
from those produced in the Higgs boson decay (``wrong pair''). This
is important for achieving precise $M_{\mathrm{H}}$ measurements
and for preventing Higgs decay mode dependence, as will be discussed
in Section \ref{sec:ModeBias}. A detailed study of the lepton pairing
algorithm can be found in\cite{Yan:2016trc}. For the Higgsstrahlung
process, the invariant mass $M_{\mathrm{l^{+}l^{-}}}$ ($\mathrm{l}$
= $\mathrm{e}$ or $\mathrm{\mu}$) of the dilepton system and recoil
mass $M_{\mathrm{rec}}$ should be close to the Z boson mass $M_{\mathrm{Z}}$=91.187
GeV \cite{1674-1137-38-9-090001} and the Higgs boson mass $M_{\mathrm{H}}$=125 GeV
(in this study), respectively. The decay modes which contain an extra
source of leptons, such as the $\mathrm{H\rightarrow ZZ^{*}}$ and
$\mathrm{H\rightarrow WW^{*}}$ modes, have a higher ratio of ``wrong
pairs''.

\begin{figure}
\centering
\includegraphics[width=\linewidth]{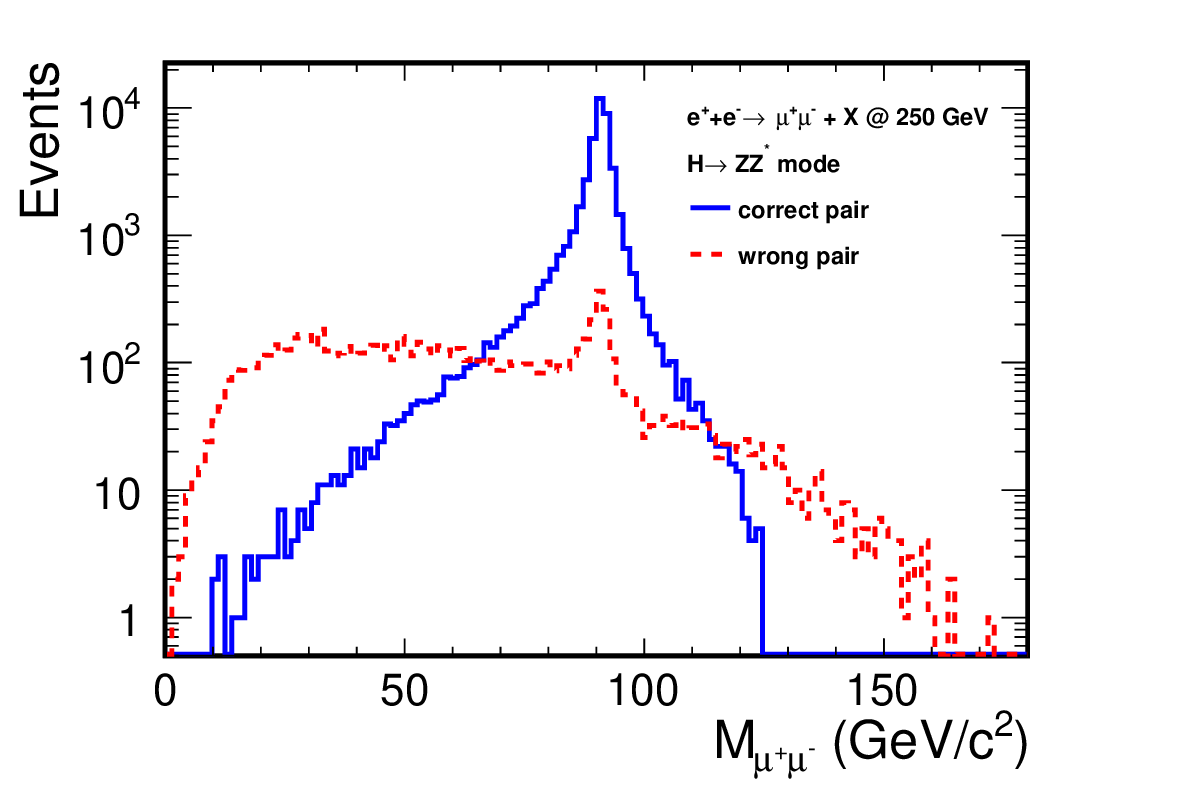} \hfill
\includegraphics[width=\linewidth]{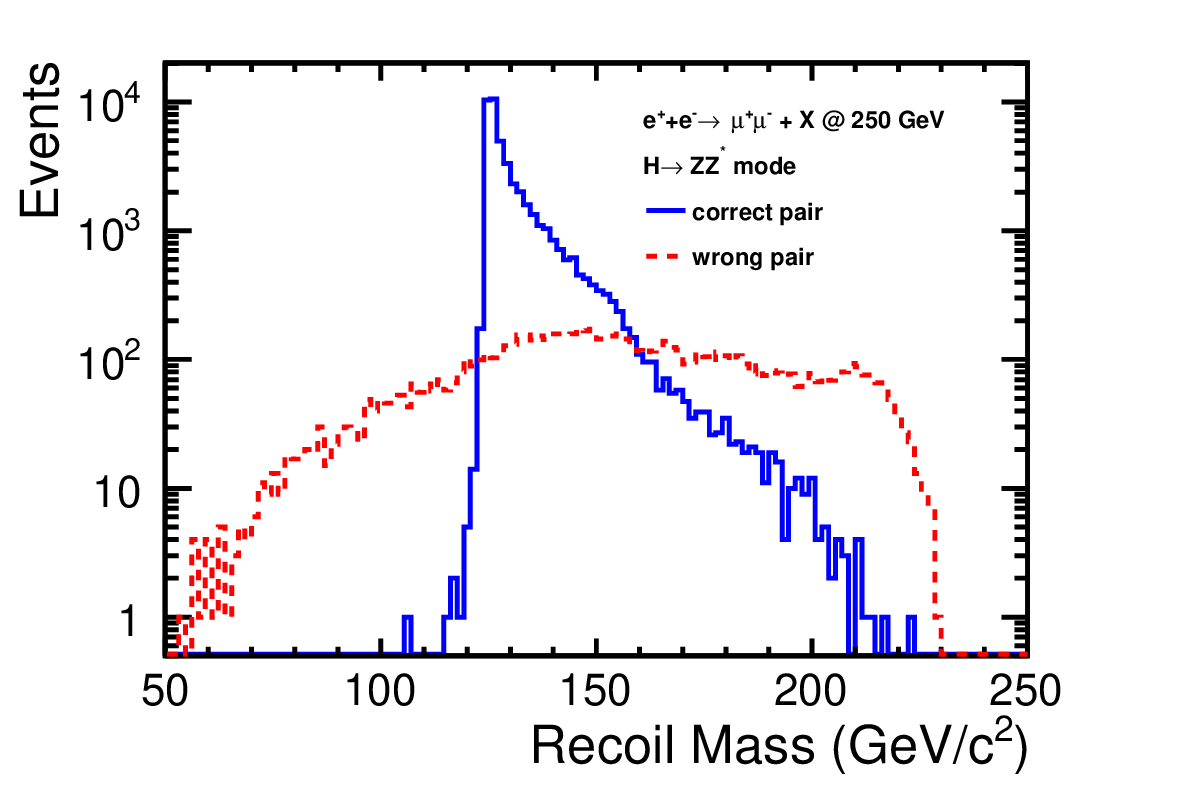}
\caption{Comparison of the distributions of $M_{\mathrm{l^{+}l^{-}}}$ (top)
and $M_{\mathrm{rec}}$ (bottom) between ``correct''
and ``wrong'' lepton pairs. This
is an example of the $\mathrm{H\rightarrow ZZ^{*}}$ decay mode in
the $\mathrm{\mu^{+}\mu^{-}H}$ channel at $\sqrt{s}$
= 250 GeV. The distributions are made using the high statistics samples
mentioned in Section \ref{sub:Signal-and-Background}.}
\label{MassCompare}
\end{figure}

The best lepton pair candidate is selected based on the following
criteria. First, the requirement $\left|M_{\mathrm{l^{+}l^{-}}}-M_{\mathrm{Z}}\right|<40(60)\:\mathrm{GeV}$
is implemented for $\mu\left(\mathrm{e}\right)$. In the case where
both leptons originate from a single Z boson produced in Higgs boson
decay, $M_{\mathrm{rec}}$ tends to deviate from $M_{\mathrm{H}}$
even if $M_{\mathrm{l^{+}l^{-}}}$ is close to $M_{\mathrm{Z}}$.
Therefore the next step is to select, taking into account both $M_{\mathrm{l^{+}l^{-}}}$
and $M_{\mathrm{rec}}$, the pair which minimizes the following $\chi^{2}$
function:

\begin{equation}
\chi^{2}\left(M_{\mathrm{l^{+}l^{-}}},M_{\mathrm{rec}}\right)=\frac{\left(M_{\mathrm{l^{+}l^{-}}}-M_{\mathrm{Z}}\right)^{2}}{\sigma_{M_{\mathrm{l^{+}l^{-}}}}^{2}}+\frac{\left(M_{\mathrm{rec}}-M_{\mathrm{H}}\right)^{2}}{\sigma_{M_{\mathrm{rec}}}^{2}}\:,\label{eq:chi2}
\end{equation}
where $\sigma_{M_{\mathrm{l^{+}l^{-}}}}$ and $\sigma_{M_{\mathrm{rec}}}$
are determined by a Gaussian fit to the distributions of $M_{\mathrm{l^{+}l^{-}}}$
and $M_{\mathrm{rec}}$ for each channel. Using the $\mathrm{H\rightarrow ZZ^{*}}$
mode in the $\mathrm{\mu^{+}\mu^{-}H}$ channel at $\sqrt{s}$=250
GeV as an example, Figure \ref{MassCompare} compares the distributions
of $M_{\mathrm{l^{+}l^{-}}}$ and $M_{\mathrm{rec}}$ between ``correct''
(solid line) and ``wrong'' (dotted
line) pairs, defined as those in which at least one lepton is from
Higgs boson decay. Here, the ``correct'' and ``wrong'' pairs are
separated using the MC truth information of the pairs selected by
the above-mentioned pairing algorithm. One can see, only in the case
of the ``correct pairs'', a clean $M_{\mathrm{l^{+}l^{-}}}$ peak
at $M_{\mathrm{Z}}$ signaling Z boson production, and a clean $M_{\mathrm{rec}}$
peak corresponding to the Higgs boson production. At $\sqrt{s}$ =
250 GeV, the efficiency of the dilepton finder described above in
finding a pair of isolated leptons is about 94\% and about 89\% for
the $\mathrm{\mu^{+}\mu^{-}H}$ and $\mathrm{e^{+}e^{-}H}$
channels, respectively. Meanwhile ``wrong pairs'' as well as the
backgrounds in Section \ref{sub:Signal-and-Background} are significantly
suppressed.

The shape of the $M_{\mathrm{rec}}$ distribution is affected by radiative
and resolution effects. The radiative effects comprise of beamstrahlung,
ISR, FSR and bremsstrahlung.
Because events are moved from the peak region of the $M_{\mathrm{rec}}$
distribution to the tail, the measurement precision is degraded. On
the other hand, resolution effects determine the peak width of the
distribution and thus the measurement uncertainties. The dominant
resolution effects are the beam energy spread induced by the accelerator
and the uncertainty of the detector response, dominated by the track
momentum resolution. Compared to these, the SM Higgs decay width of
about 4 MeV is negligible. While ISR and FSR are irreducible physical
effects, beamstrahlung, bremsstrahlung, and resolution effects can
be mitigated by optimization in the design of accelerator and detector.

\subsection{Recovery of Bremsstrahlung and FSR Photons}
\label{sub:Recovery-of-Bremsstrahlung}

\begin{figure}
\centering
\includegraphics[width=.425\textwidth]{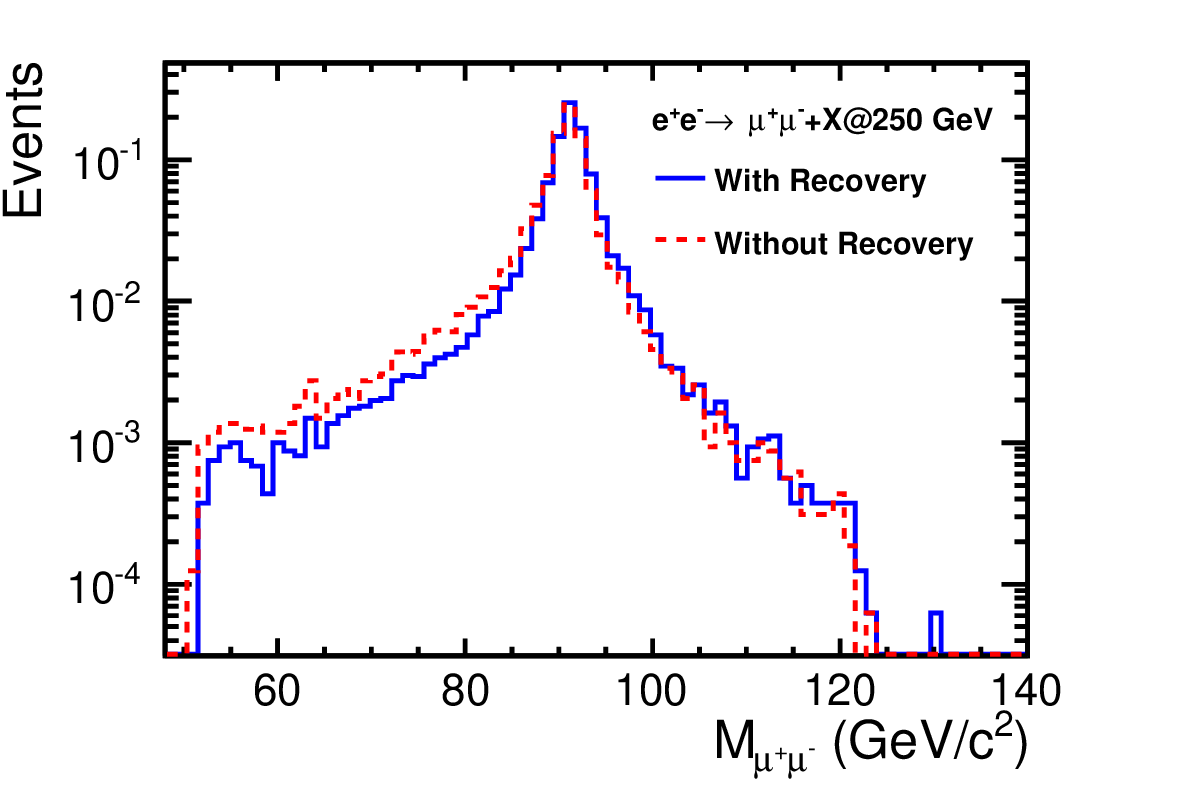} \hfill
\includegraphics[width=.425\textwidth]{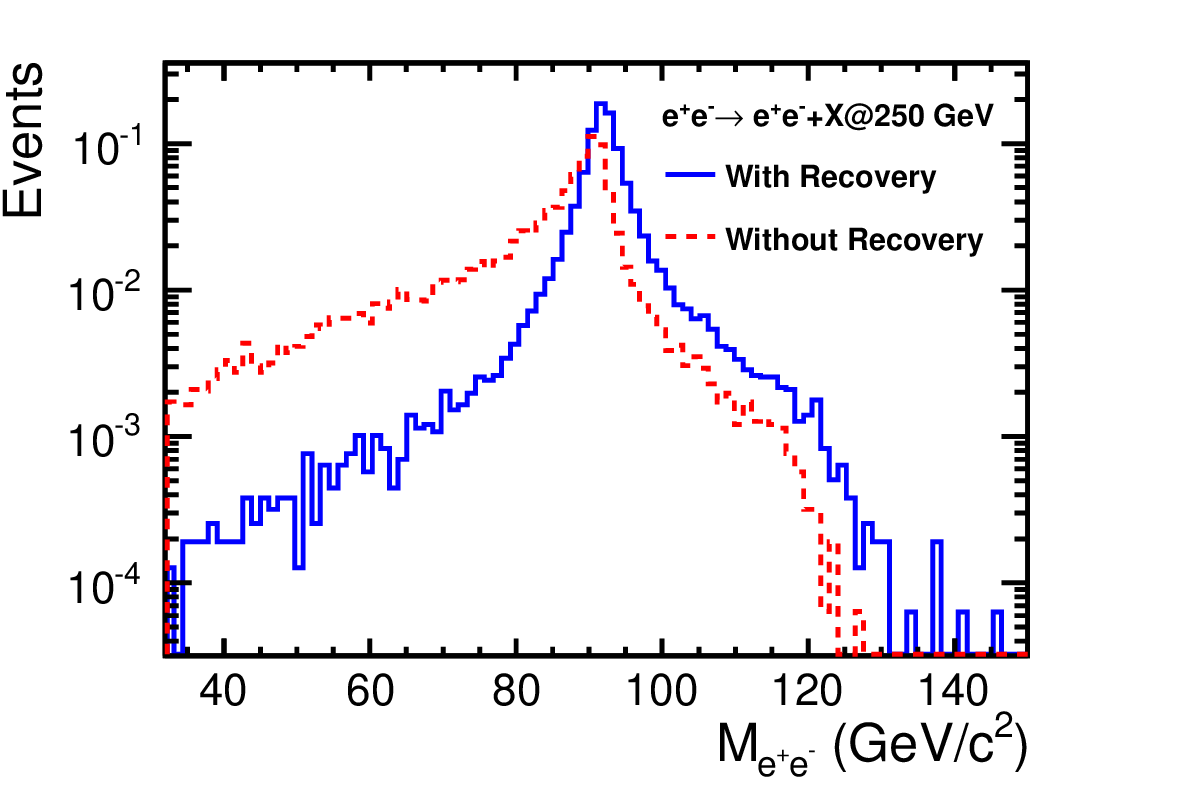} \\
\includegraphics[width=.425\textwidth]{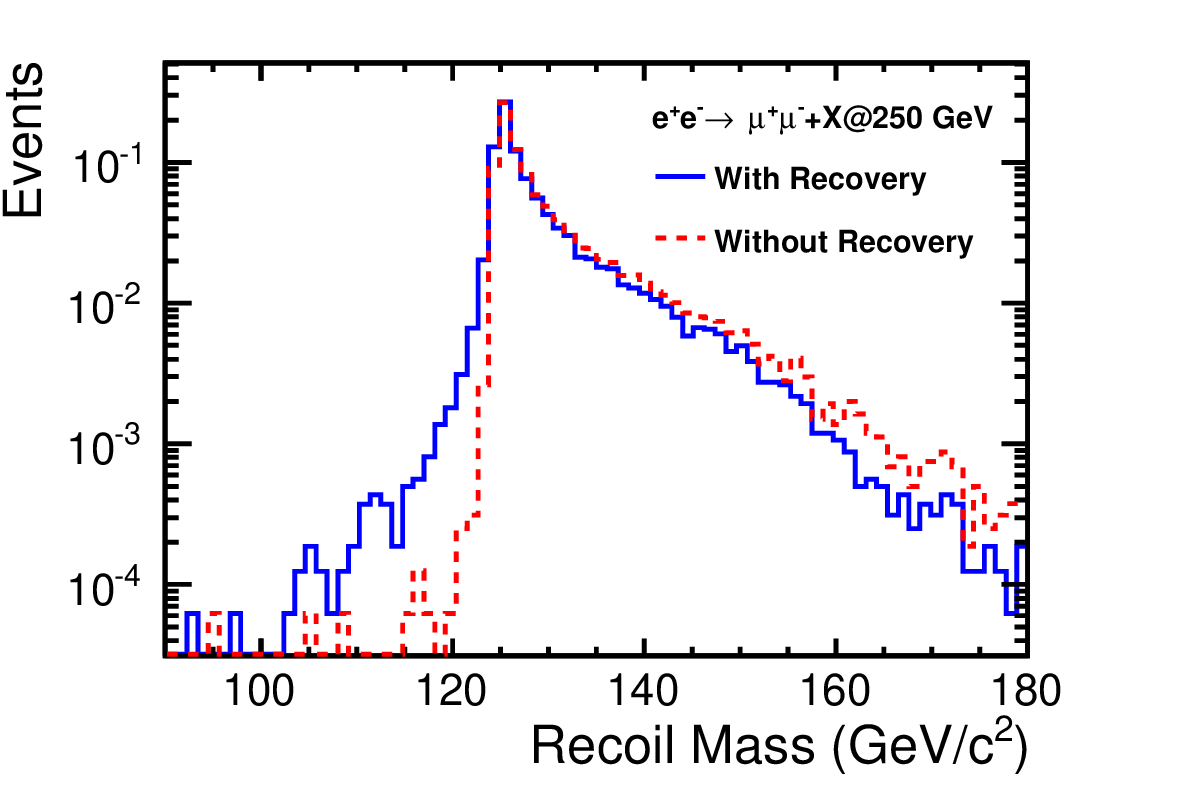} \hfill
\includegraphics[width=.425\textwidth]{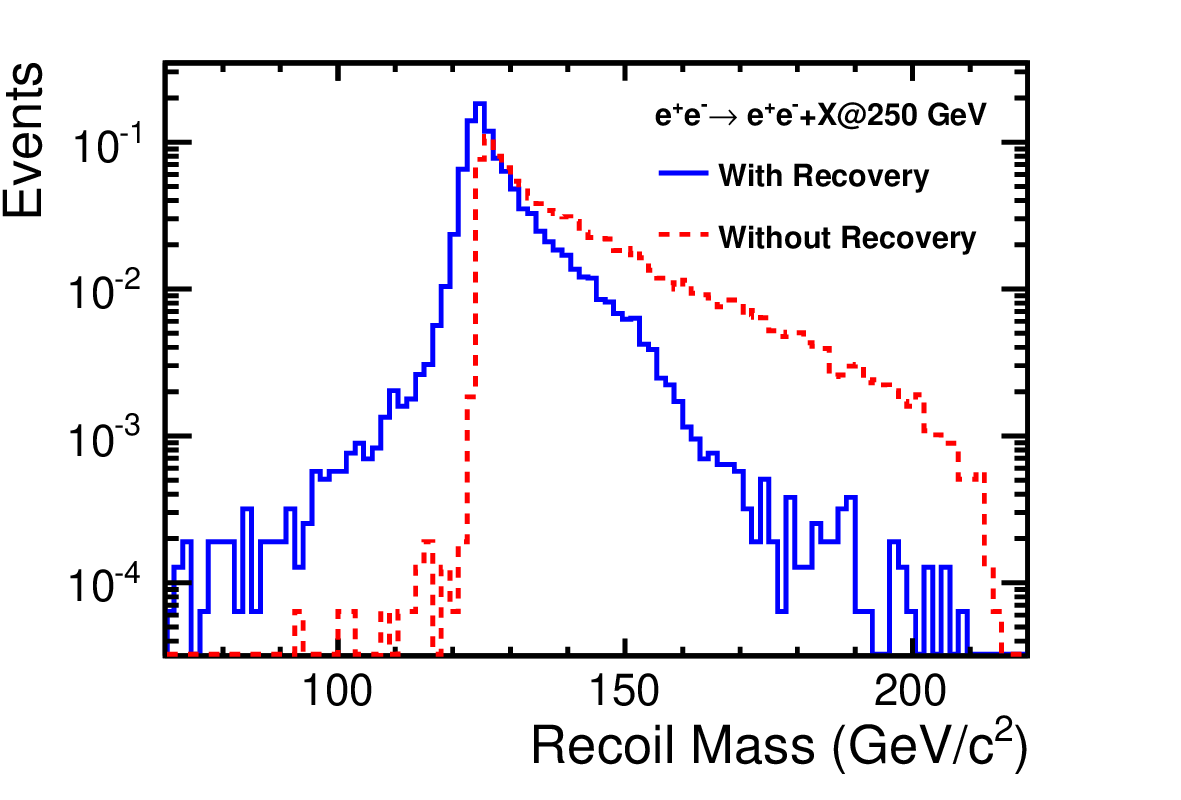}
\caption{Comparison of the $M_{\mathrm{l^{+}l^{-}}}$ (two topmost) and $M_{\mathrm{rec}}$
(two bottommost) spectra between the cases with (blue) and without (red) bremsstrahlung/FSR
recovery for $\sqrt{s}$ = 250 GeV. The two bottommost plots show the $\mathrm{\mu^{+}\mu^{-}H}$
and $\mathrm{e^{+}e^{-}H}$ channels, respectively. The histograms
are normalized to unit area.}
\label{Recovery}
\end{figure}

The bremsstrahlung and FSR of the final state leptons degrade measurement
precision of $\sigma_{\mathrm{ZH}}$ and $M_{\mathrm{H}}$, particularly
for the $\mathrm{e^{+}e^{-}H}$ channel. The $M_{\mathrm{rec}}$ distribution
of the $\mathrm{e^{+}e^{-}H}$ channel has a broader peak and longer
tail to lower values than the $\mathrm{\mu^{+}\mu^{-}H}$
channel. The recovery of bremsstrahlung and FSR photons is implemented
for both $\mathrm{\mu^{+}\mu^{-}H}$ and $\mathrm{e^{+}e^{-}H}$
channels. A bremsstrahlung/FSR photon is identified using its polar
angle with respect to the final state lepton; if the cosine of the
polar angle exceeds 0.99, the photon four momentum is combined with
that of the lepton. Figure \ref{Recovery} compares the reconstructed
$M_{\mathrm{l^{+}l^{-}}}$ and $M_{\mathrm{rec}}$ spectra before
(dotted line) and after (solid line) bremsstrahlung/FSR recovery for
$\sqrt{s}$ = 250 GeV. It can be seen that the recovery process pushes
the events at the lower end of the $M_{\mathrm{l^{+}l^{-}}}$ spectrum
(corresponding to the tail in the higher region of the $M_{\mathrm{rec}}$
spectrum) back to the peak.

\subsection{Background Rejection}
\label{sub:Background-Rejection}

\begin{figure}
\centering
\includegraphics[width=\linewidth]{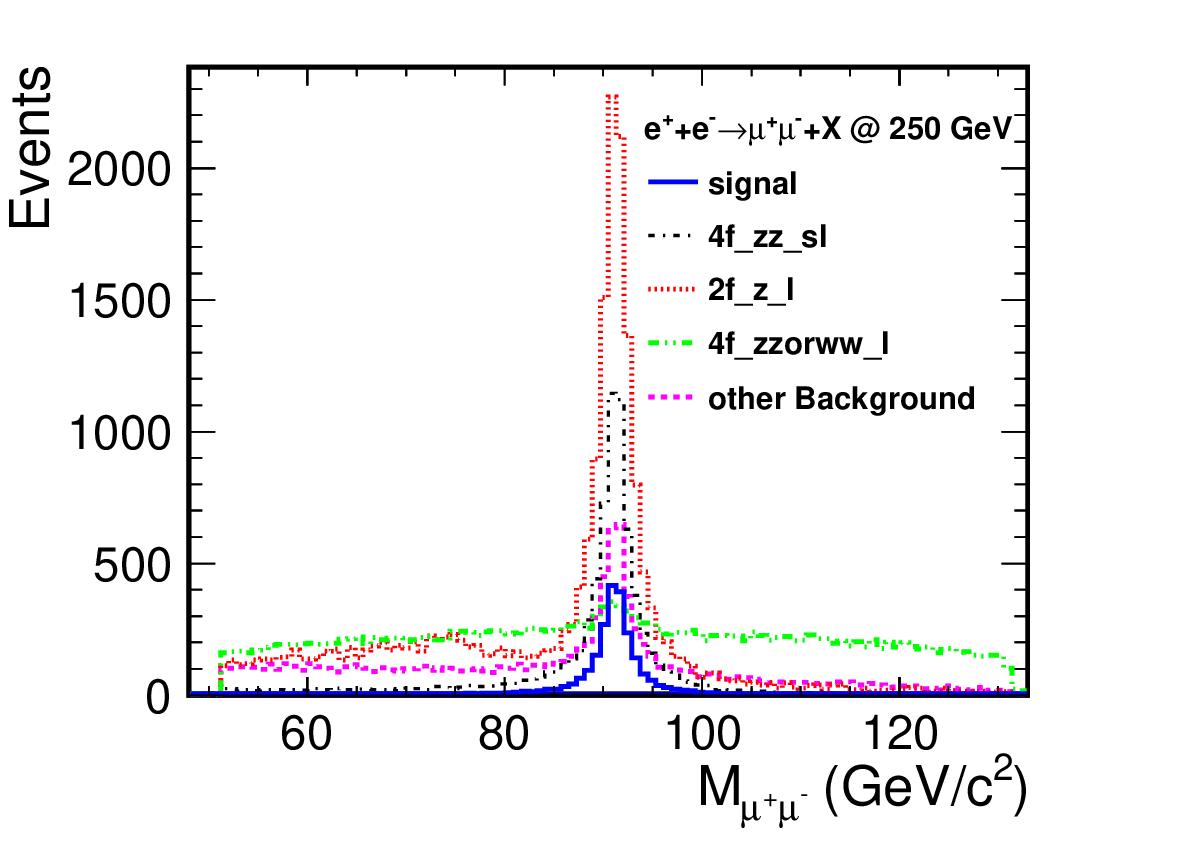} \hfill
\includegraphics[width=\linewidth]{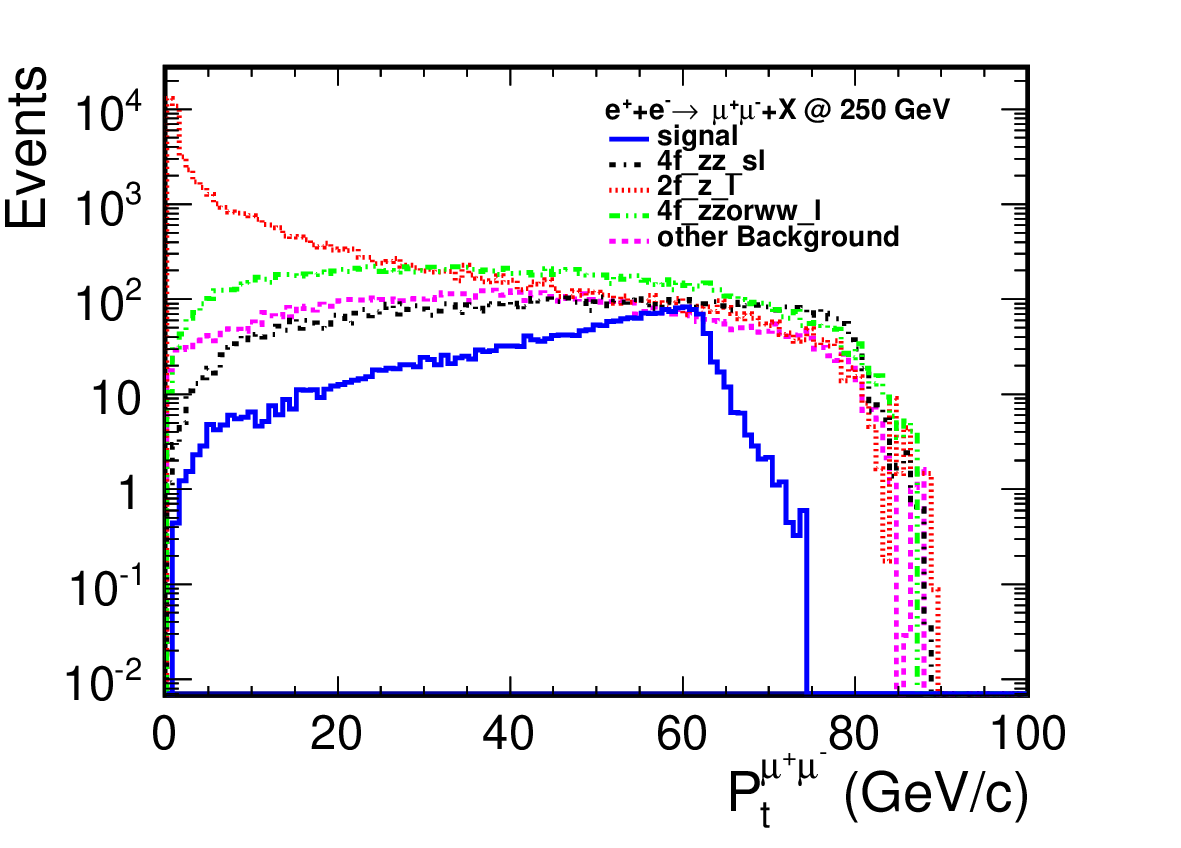} \hfill
\includegraphics[width=\linewidth]{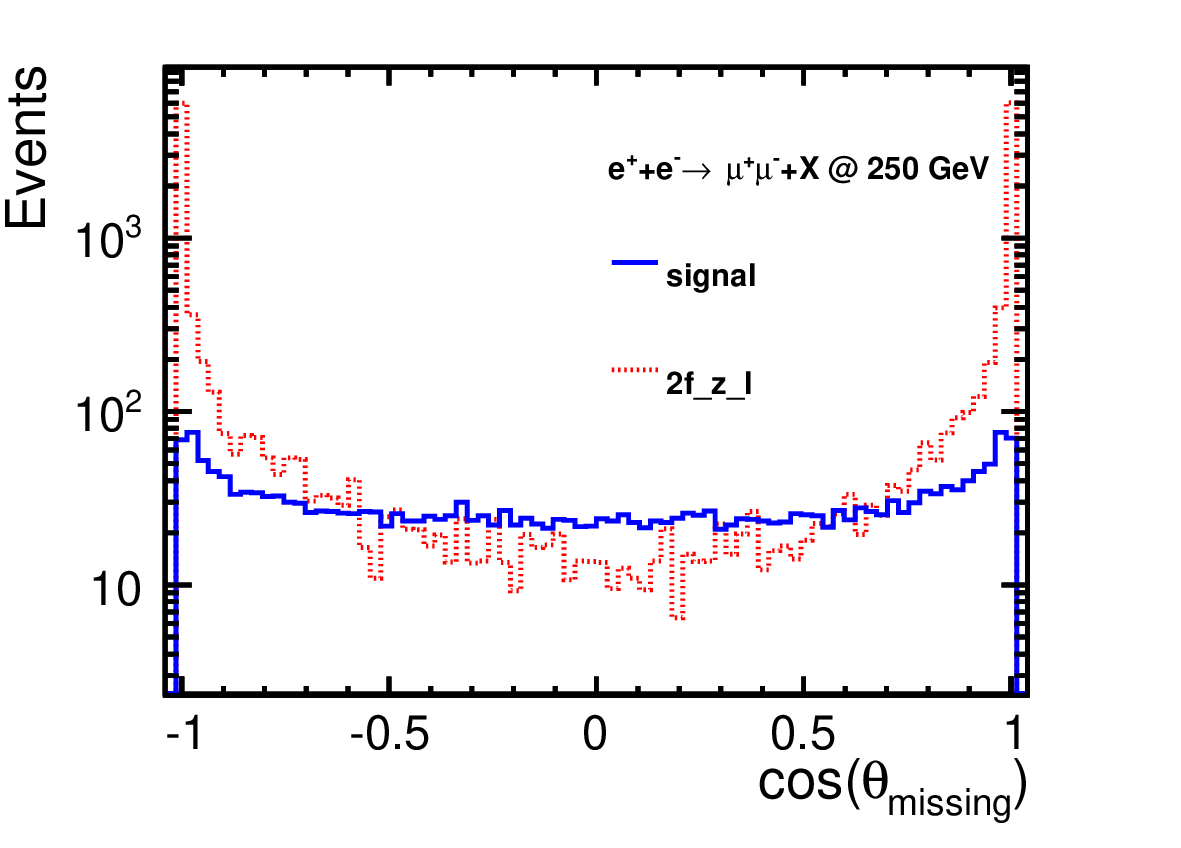}

\caption{(top) The $M_{\mu^{+}\mu^{-}}$ distributions of signal and the major
background processes, after a loose precut on $M_{\mathrm{rec}}$.
(center) The $p_{\mathrm{\mathrm{T}}}^{\mu^{+}\mu^{-}}$ distributions
of signal and the major background processes, after a loose precut
on $M_{\mathrm{rec}}$ and a cut on $M_{\mu^{+}\mu^{-}}$. (bottom)
The $\mathrm{cos\left(\theta_{missing}\right)}$ distributions of
signal and 2-fermion background, after a loose precut on $M_{\mathrm{rec}}$
and cuts have been applied on $M_{\mu^{+}\mu^{-}}$ and $p_{\mathrm{\mathrm{T}}}^{\mu^{+}\mu^{-}}$.
\label{fig:-Minv}}
\end{figure}

\begin{figure*}
\centering
\includegraphics[width=.32\textwidth]{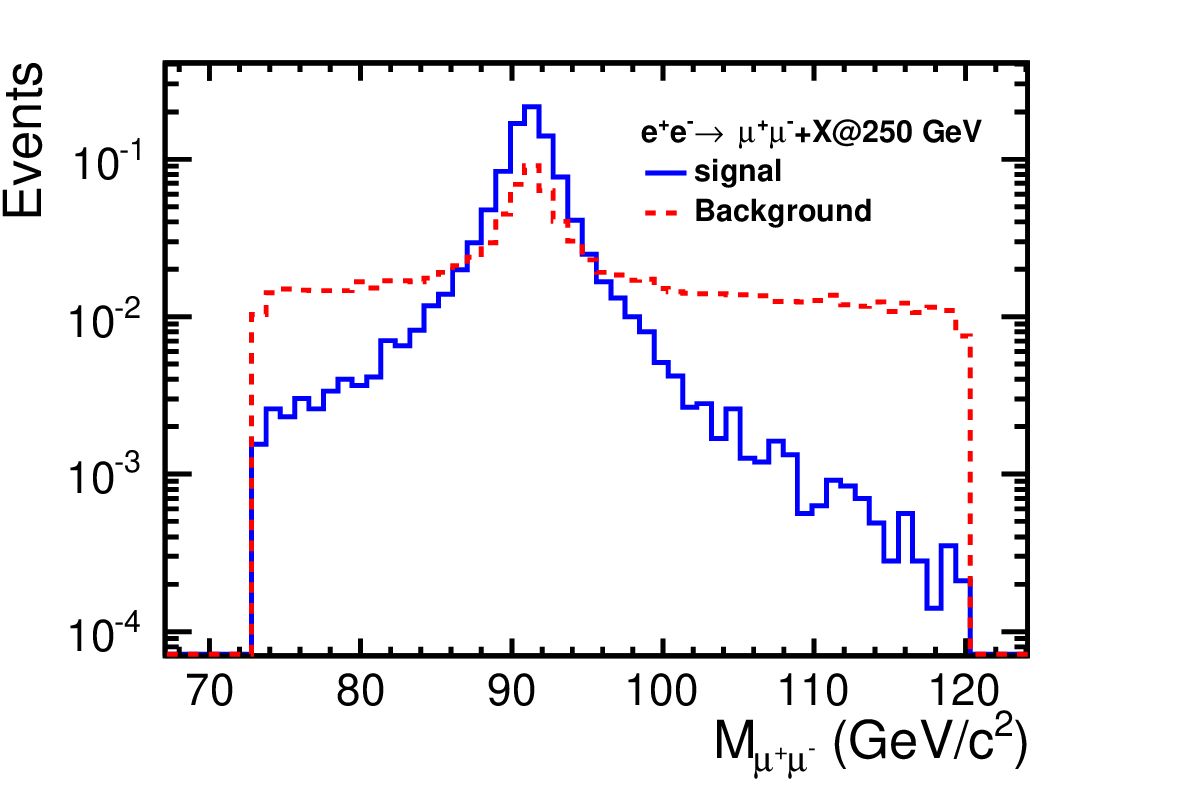} \hfill
\includegraphics[width=.32\textwidth]{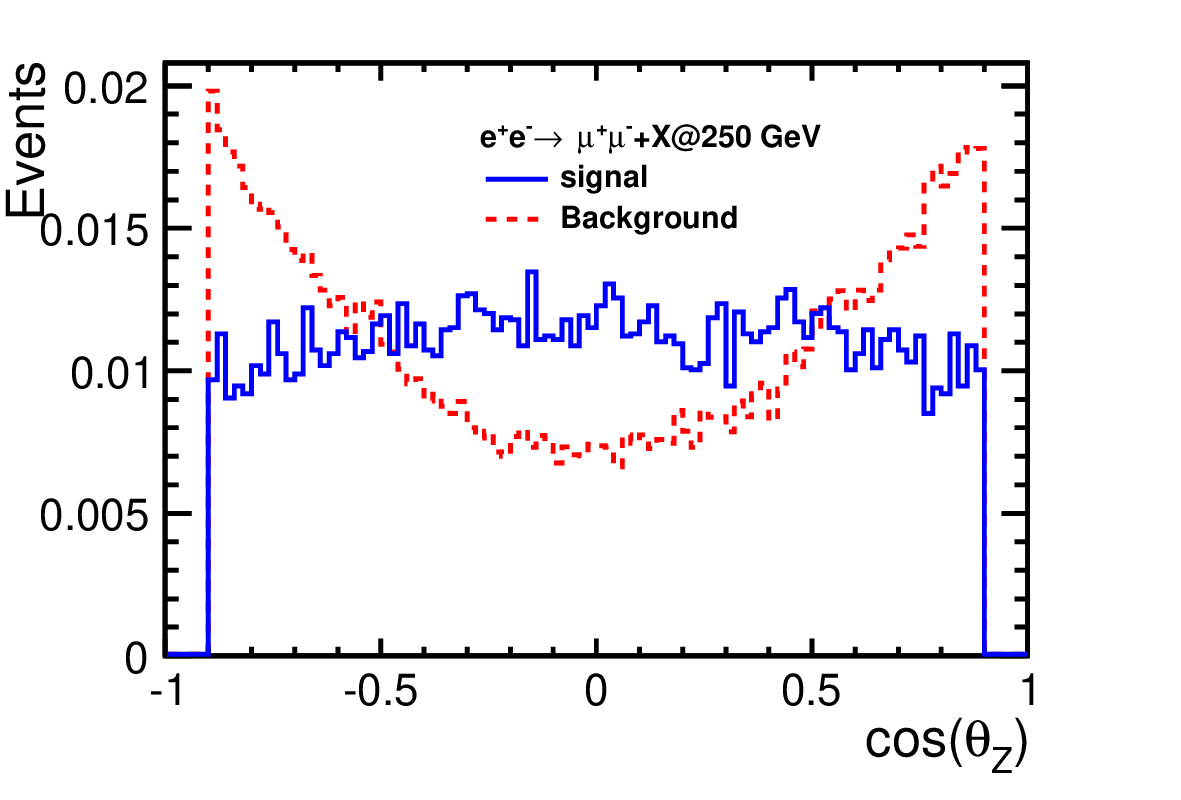} \hfill
\includegraphics[width=.32\textwidth]{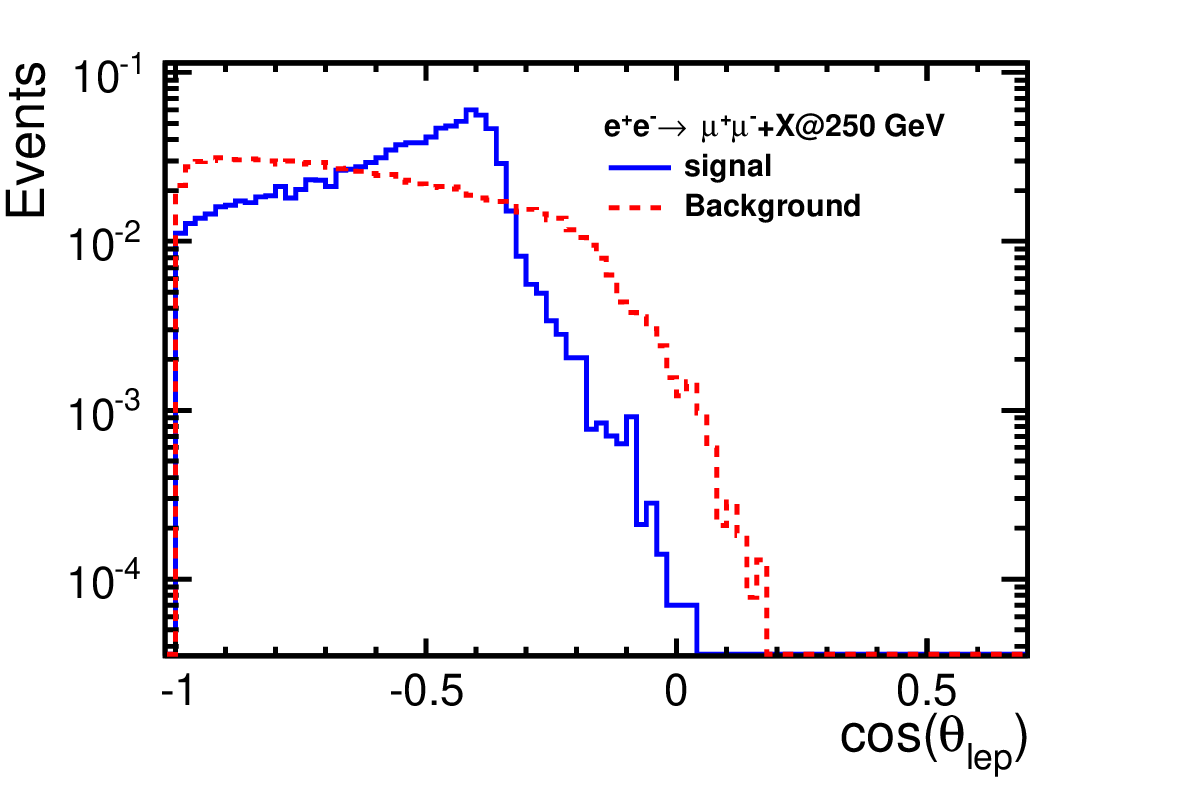}

\includegraphics[width=.32\textwidth]{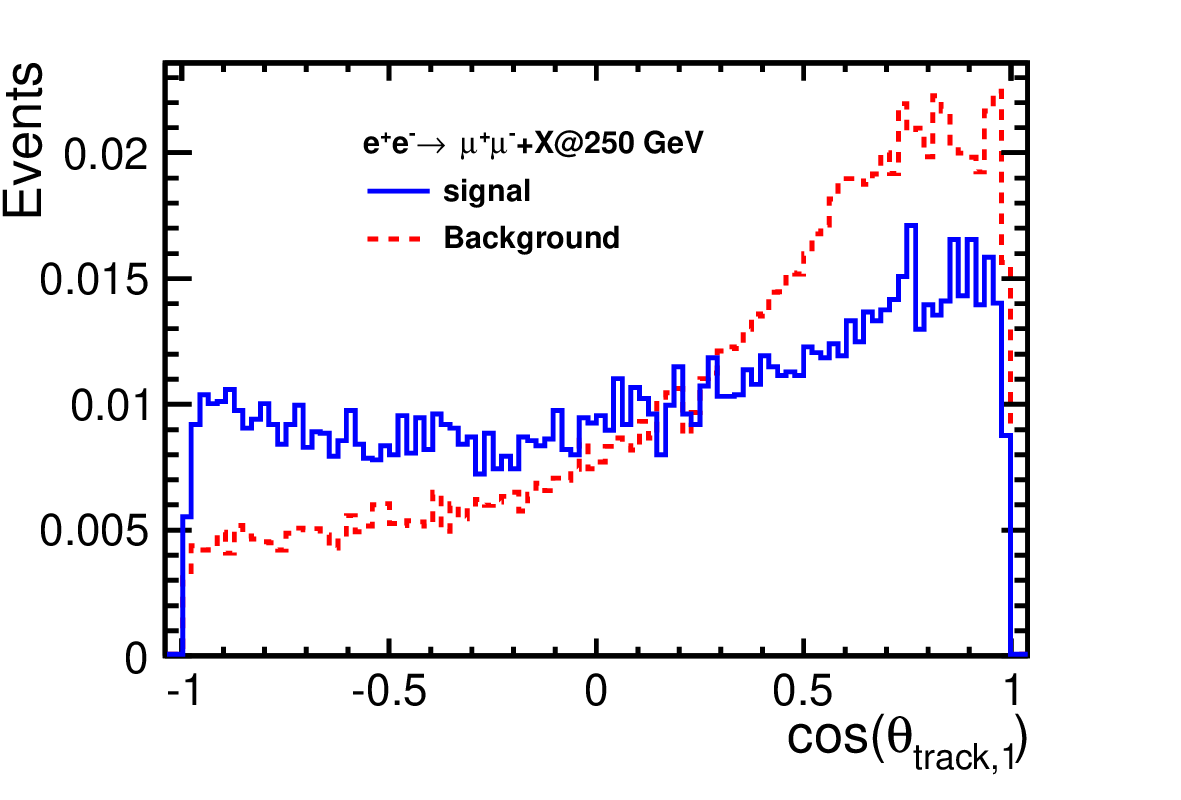} \hfill
\includegraphics[width=.32\textwidth]{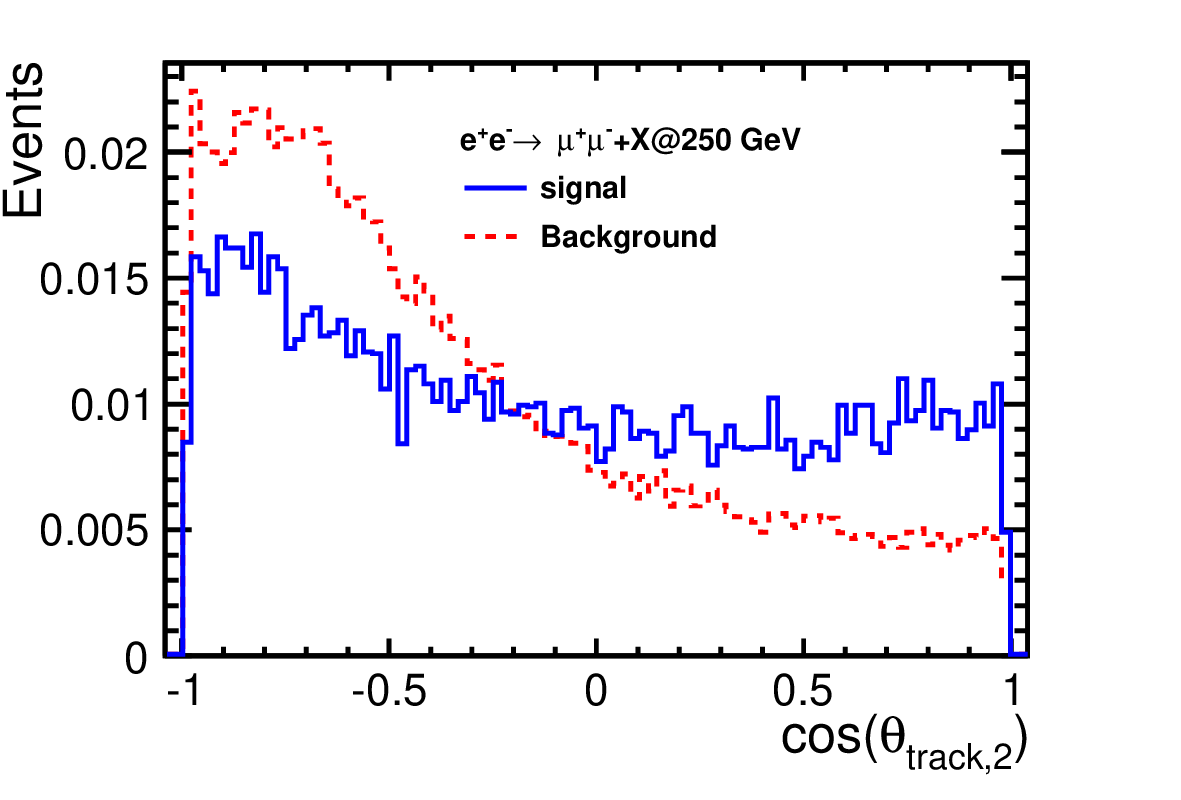} \hfill
\includegraphics[width=.32\textwidth]{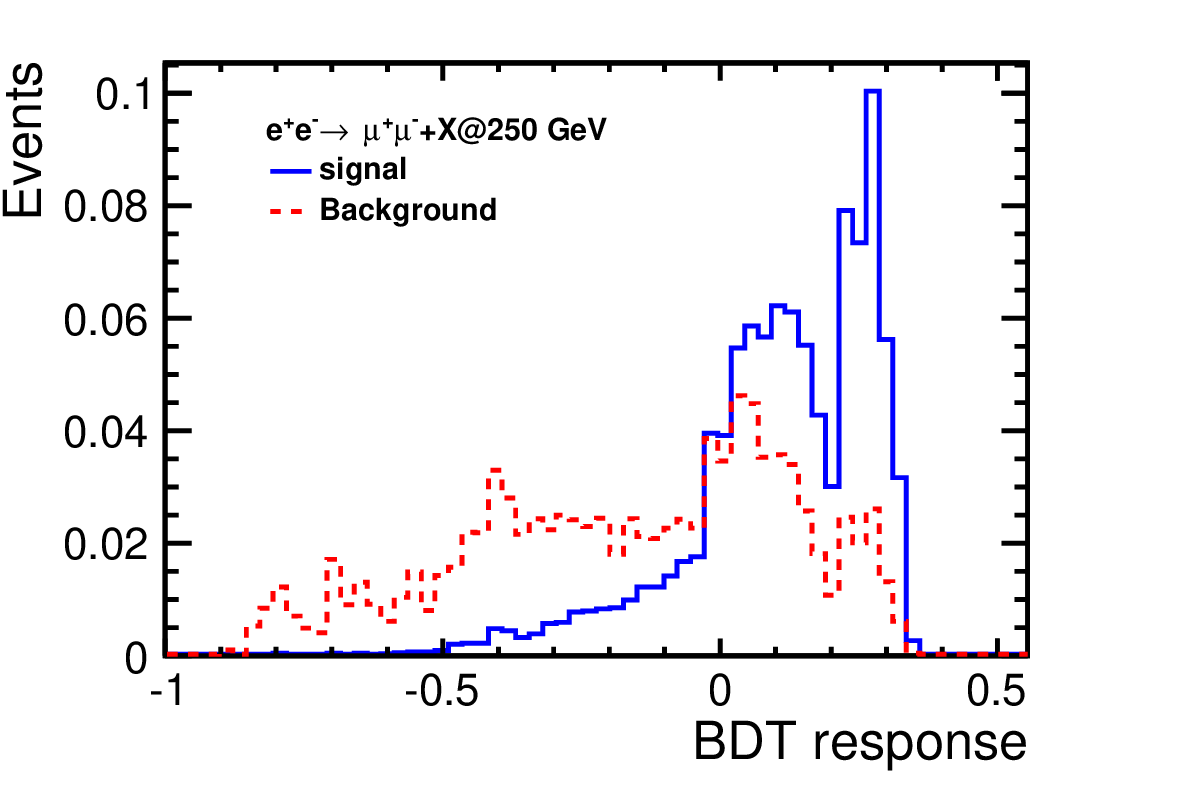}

\caption{The distributions of the variables $M_{\mu^{+}\mu^{-}}$ , $\mathrm{cos\left(\theta_{Z}\right)}$,
$\mathrm{cos\left(\theta_{lep}\right)}$, $\mathrm{cos\left(\theta_{track,1}\right)}$,
and $\mathrm{cos\left(\theta_{track,2}\right)}$ used for the training
in the multi-variate analysis, as well as the distribution of the
BDT response, shown here for the signal and background in the case
of the $\mathrm{\mu^{+}\mu^{-}H}$ channel at $\sqrt{s}$=250
GeV, after a loose precut on $M_{\mathrm{rec}}$ and cuts have been
applied on $M_{\mu^{+}\mu^{-}}$, $p_{\mathrm{\mathrm{T}}}^{\mu^{+}\mu^{-}}$,
and $\mathrm{cos\left(\theta_{missing}\right)}$. The histograms are
normalized.}
\label{fig:likelihood-Var}
\end{figure*}

\begin{figure}
\centering
\includegraphics[width=\linewidth]{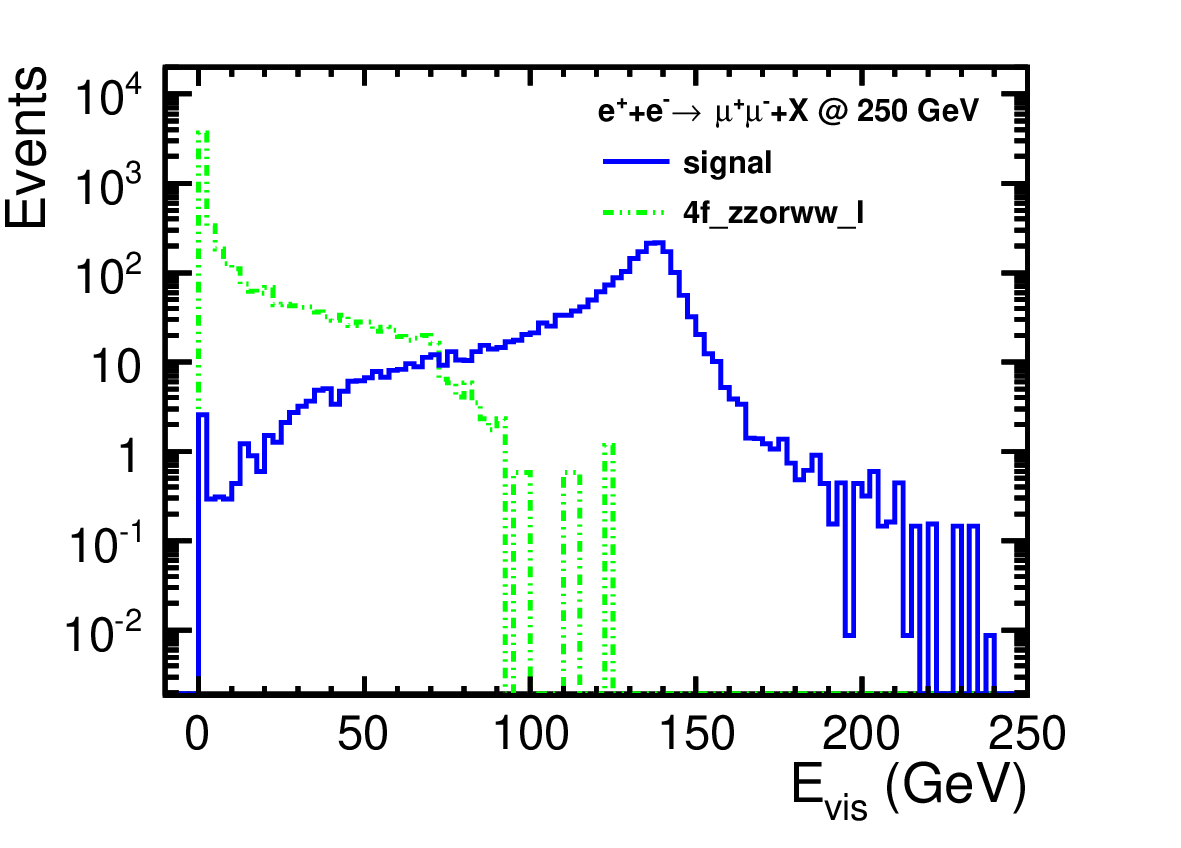}
\caption{The distributions of $E_{\mathrm{vis}}$ (after excluding the dilepton
energy) of the signal and the 4f\_zzorww\_l processes, after a loose
precut on $M_{\mathrm{rec}}$ and cuts have been applied on $M_{\mu^{+}\mu^{-}}$,
$p_{\mathrm{\mathrm{T}}}^{\mu^{+}\mu^{-}}$, $\mathrm{cos\left(\theta_{missing}\right)}$,
and the BDT response of the MVA analysis.}
\label{fig:Evis}
\end{figure}

\begin{figure}
\centering
\includegraphics[width=\linewidth]{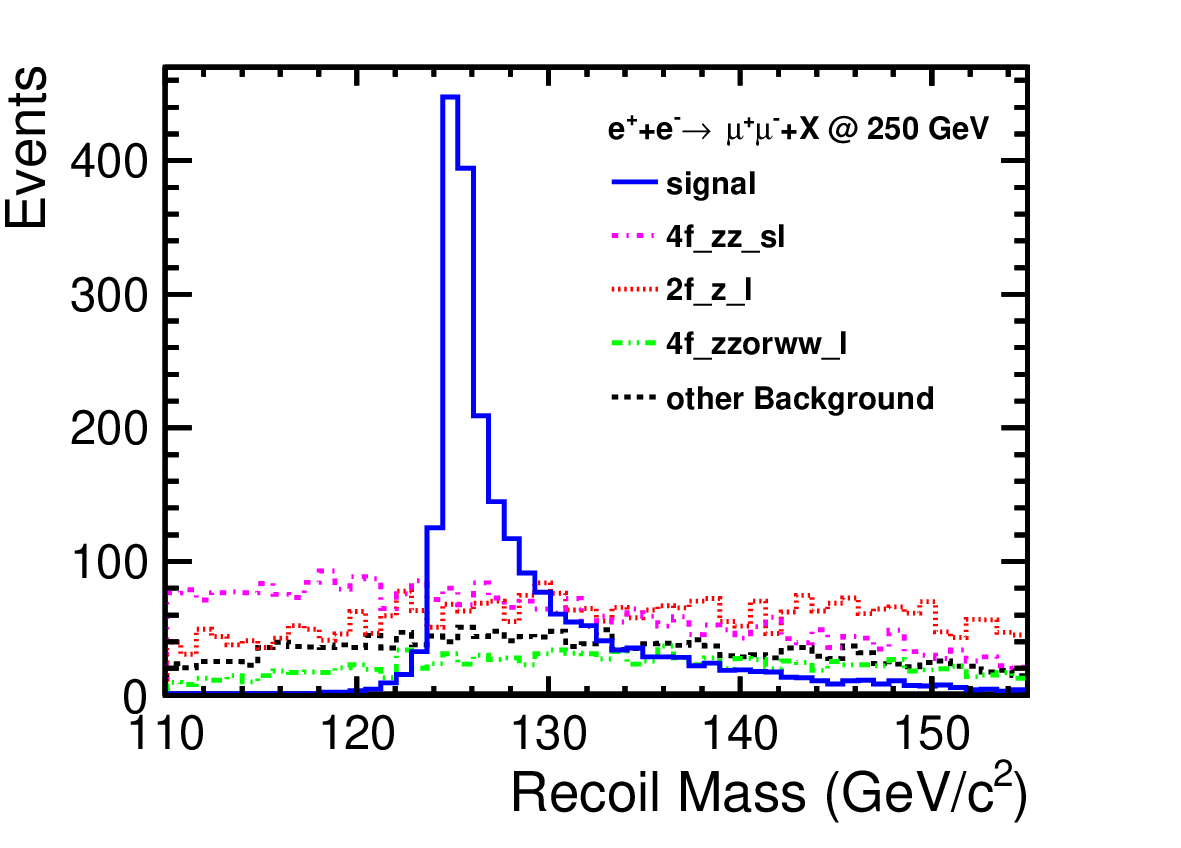} \hfill
\includegraphics[width=\linewidth]{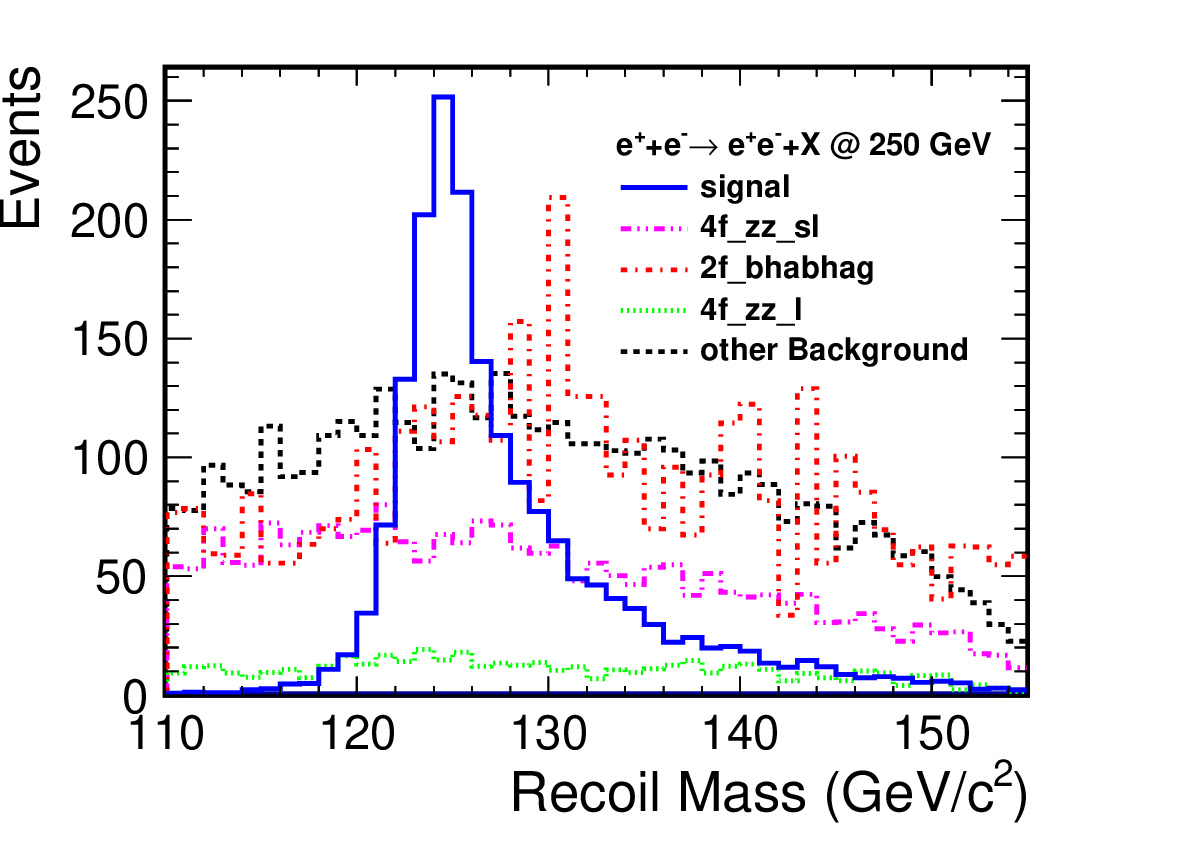}
\caption{The histograms of the recoil mass of the signal and the major residual
background processes left in a wide window around the signal $M_{\mathrm{rec}}$
peak, shown here for the $\mathrm{\mu^{+}\mu^{-}H}$ (top)
and $\mathrm{e^{+}e^{-}H}$
(bottom) channels at $\sqrt{s}$=250 GeV, after all cuts described
in the main text have been applied.}
\label{fig:stk}
\end{figure}

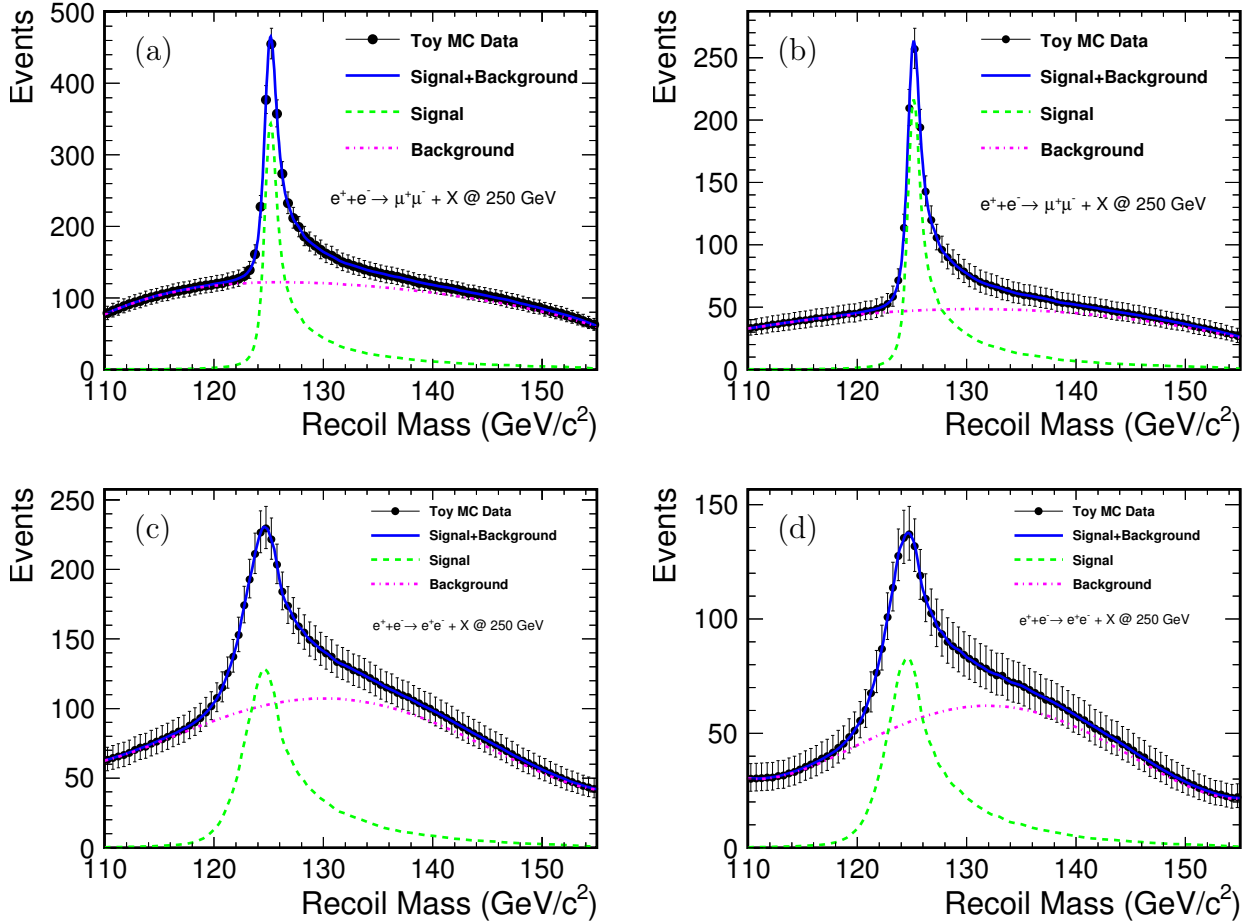
\begin{figure*}
\centering
    \begin{picture}(240,180)
        \put(0,0){
        \includegraphics[width=.5\textwidth]{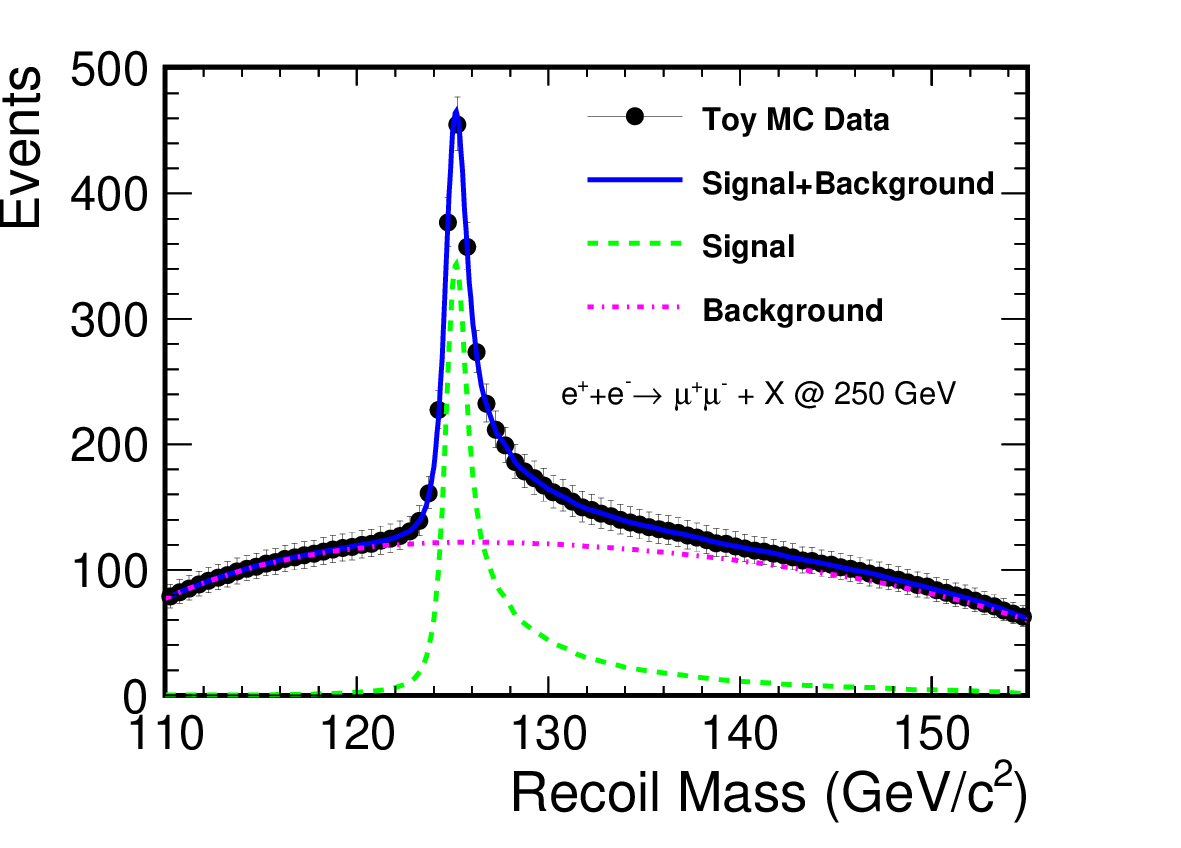}
        }
        \put(50,150){\large(a)}
    \end{picture}
    \begin{picture}(240,180)
        \put(0,0){
        \includegraphics[width=.5\textwidth]{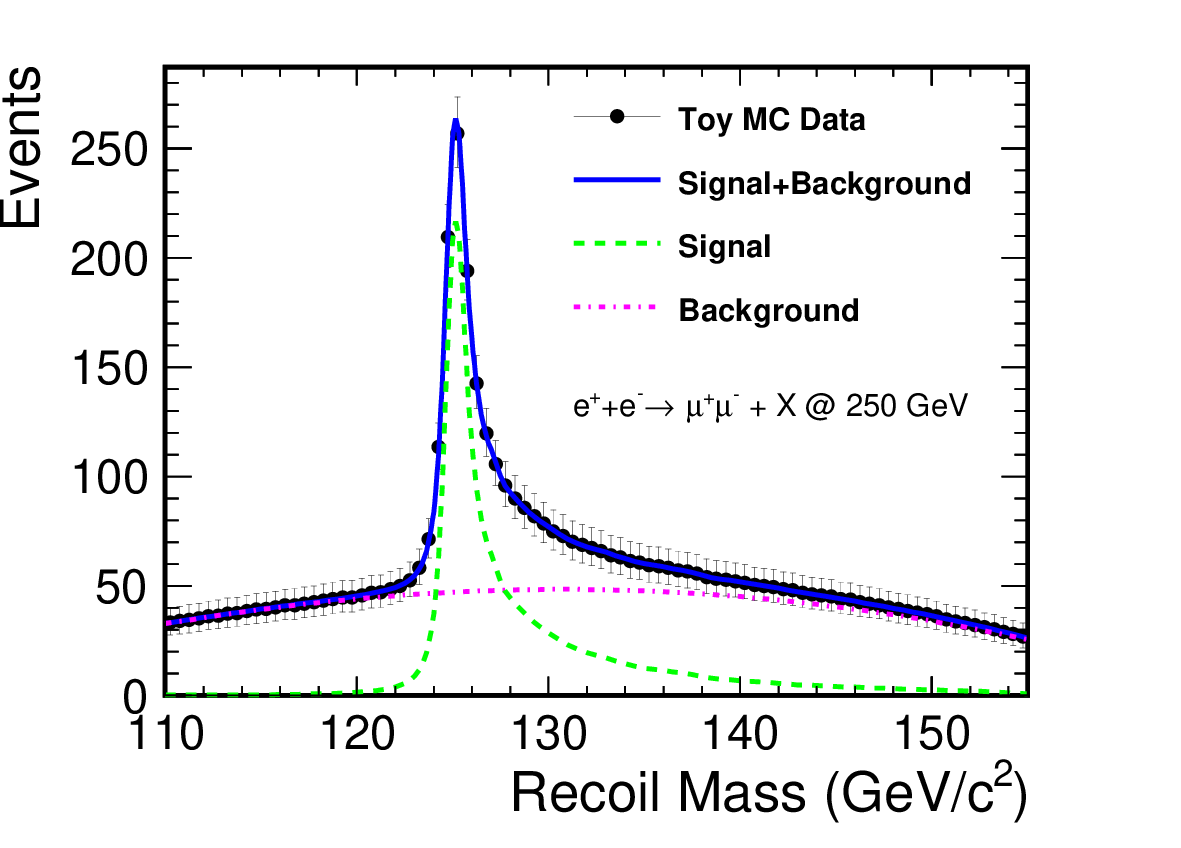}
        }
        \put(50,150){\large(b)}
    \end{picture} \\
    \begin{picture}(240,180)
        \put(0,0){
        \includegraphics[width=.5\textwidth]{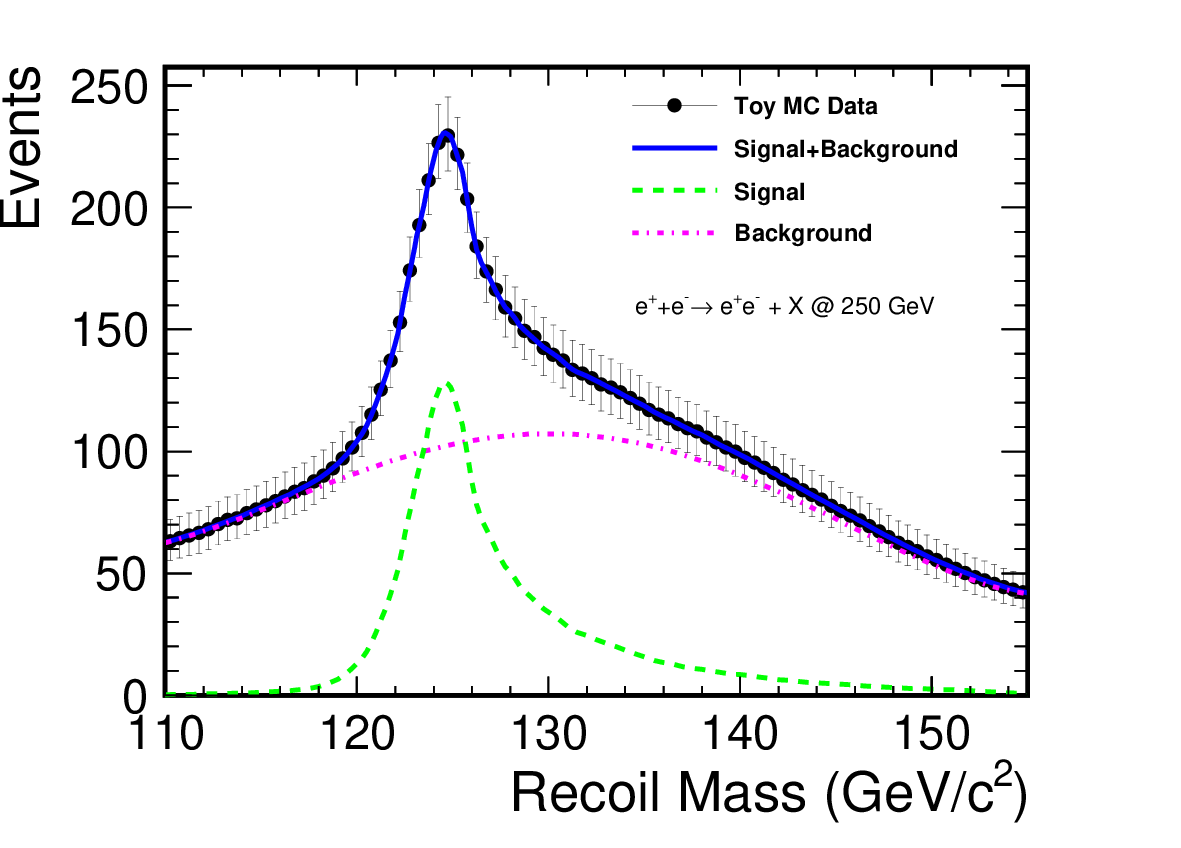}
        }
        \put(50,150){\large(c)}
    \end{picture}
    \begin{picture}(240,180)
        \put(0,0){
        \includegraphics[width=.5\textwidth]{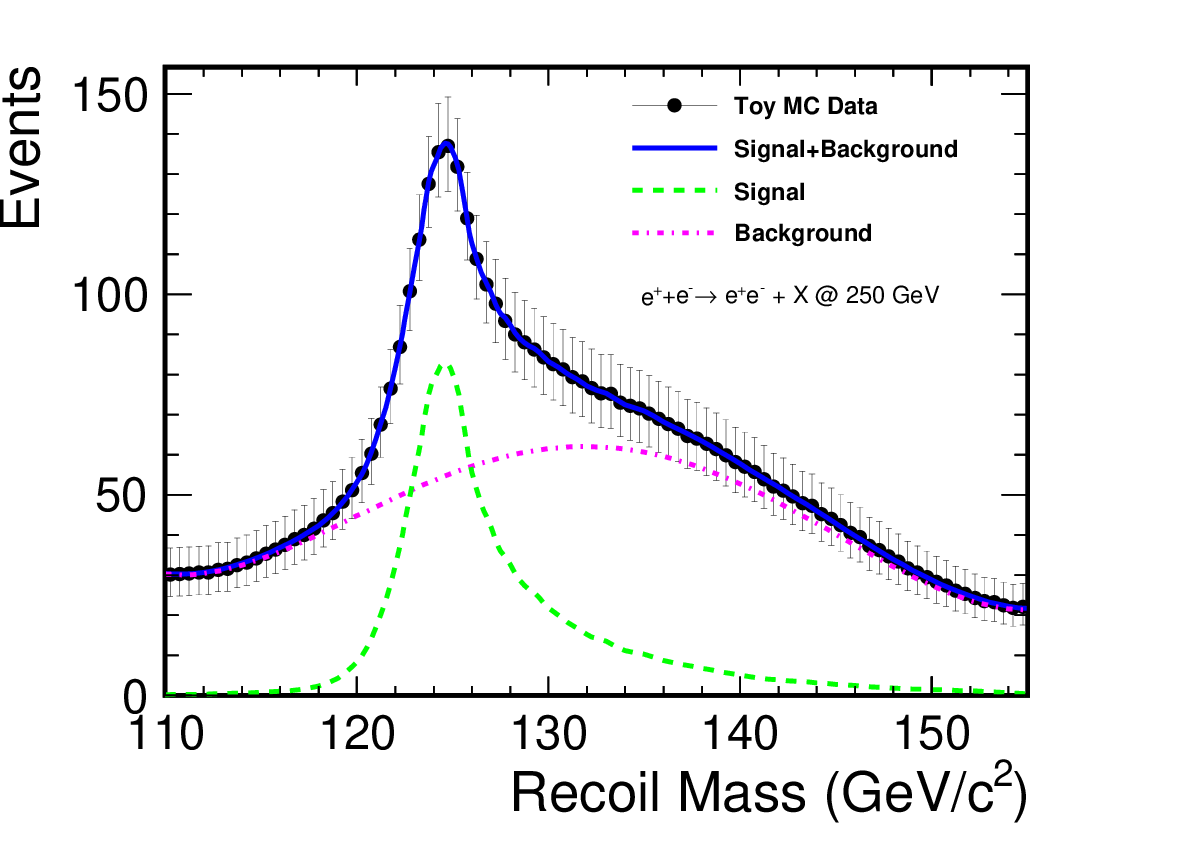}
        }
        \put(50,150){\large(d)}
    \end{picture}

\caption{The recoil mass spectra of events in the signal region 110-155 GeV
at $\sqrt{s}$ = 250 GeV: (a) $\mathrm{\mu^{+}\mu^{-}H}$,
$\mathrm{e_{L}^{-}e_{R}^{+}}$ (b) $\mathrm{\mu^{+}\mu^{-}H}$,
$\mathrm{e_{R}^{-}e_{L}^{+}}$ (c) $\mathrm{e^{+}e^{-}H}$,
$\mathrm{e_{L}^{-}e_{R}^{+}}$ (d) $\mathrm{e^{+}e^{-}H}$,
$\mathrm{e_{R}^{-}e_{L}^{+}}$.
The fitting functions used for the extraction of $\sigma_{\mathrm{ZH}}$
and $M_{\mathrm{H}}$ (see Section \ref{sub:Method-of-Fitting}) are
superimposed. The black markers are the Monte Carlo (MC) data points,
the green, magenta, and blue lines indicate the fitted function for
signal, background, and the combination of signal and background,
respectively.}
\label{fig:rec250}
\end{figure*}

\begin{figure}
\centering
\includegraphics[width=\linewidth]{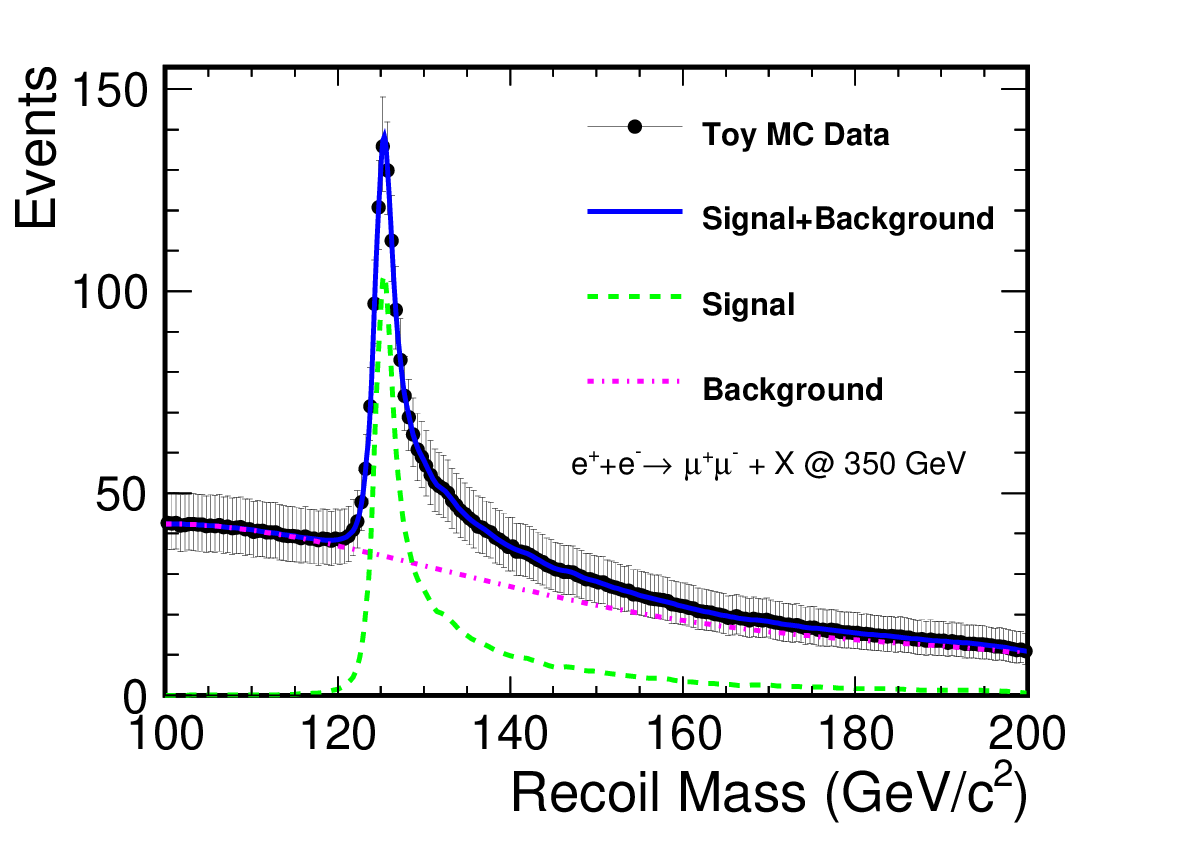}
\includegraphics[width=\linewidth]{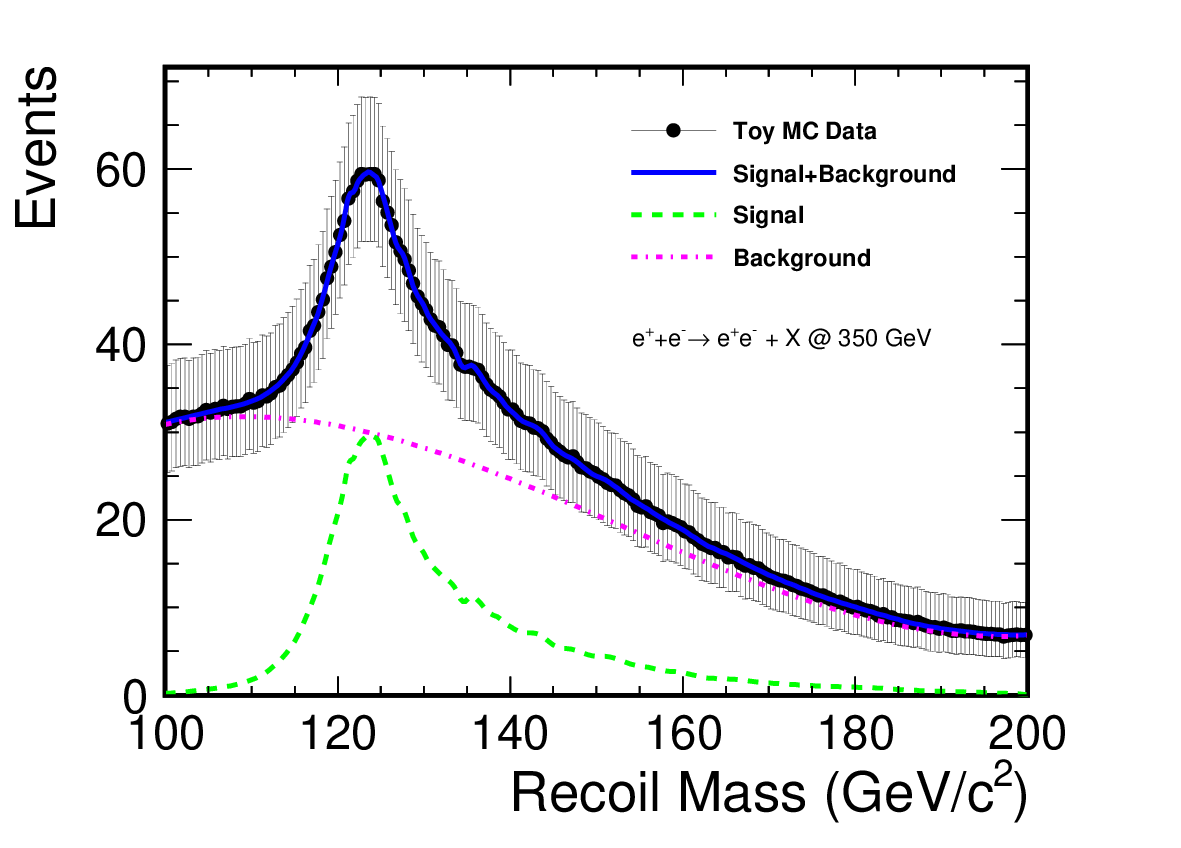}

\caption{The recoil mass spectra of events in the signal region 100-200 GeV
for $\sqrt{s}$ = 350 GeV. Top: $\mathrm{\mu^{+}\mu^{-}H}$,
$\mathrm{e_{L}^{-}e_{R}^{+}}$ Bottom $\mathrm{e^{+}e^{-}H}$,
$\mathrm{e_{L}^{-}e_{R}^{+}}$. The legend
is same as in Figure \ref{fig:rec250}.}
\label{fig:rec350}
\end{figure}

\begin{figure}
\centering
    \includegraphics[width=\linewidth]{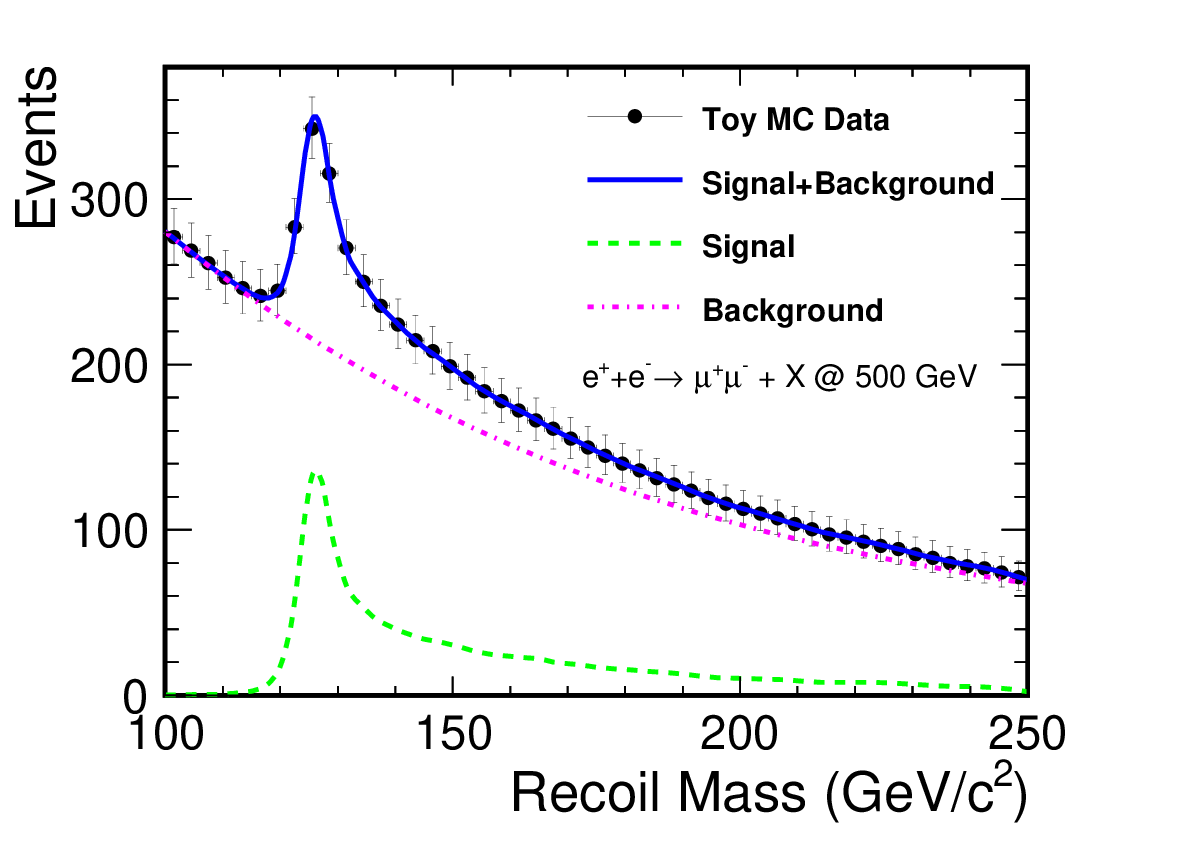}
    \includegraphics[width=\linewidth]{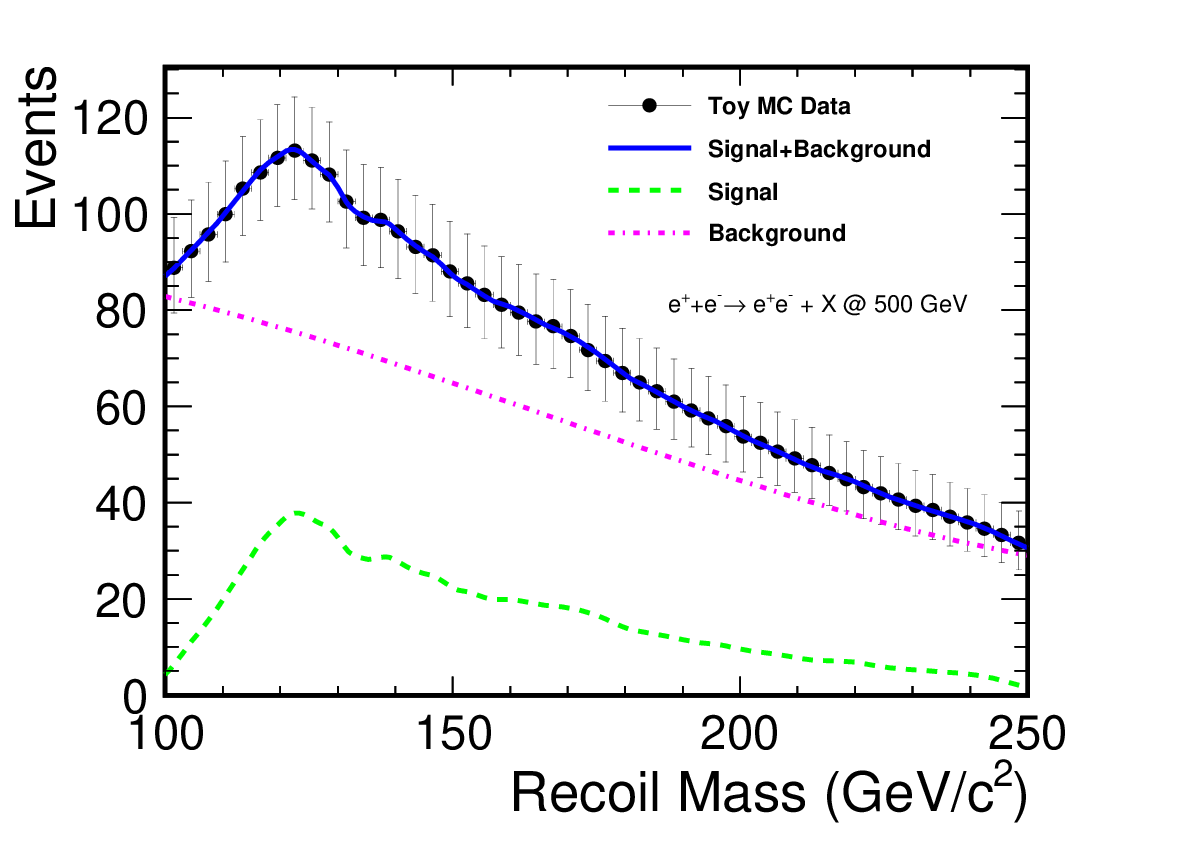}

\caption{The recoil mass spectra of events in the signal region 100-250 GeV
for $\sqrt{s}$ = 500 GeV. (a) $\mathrm{\mu^{+}\mu^{-}H}$,
$\mathrm{e_{L}^{-}e_{R}^{+}}$ (b) $\mathrm{e^{+}e^{-}H}$,
$\mathrm{e_{L}^{-}e_{R}^{+}}$. The legend
is same as in Figure \ref{fig:rec250}.\label{fig:rec500}}
\end{figure}

After the signal selection process, background events are rejected
by applying cuts on various kinematic properties. While the cut values
are adjusted for each center-of-mass energy, the overall strategies
are similar. Unless specified otherwise, the plots in this section
are shown for the case of the $\mathrm{\mu^{+}\mu^{-}H}$
channel and $\mathrm{e_{L}^{-}e_{R}^{+}}$
polarization at $\sqrt{s}$=250 GeV. In these plots, background with ZZ intermediate states and two pairs of
$\mu\mu$ / $\tau\tau$ (a pair of $\mu\mu$ / $\tau\tau$ and a pair
of quarks) is denoted with 4f\_zz\_l(sl), background with final
states of $\mu\mu$ / $\tau\tau$ and $\mathrm{ee}$ is denoted with 2f\_z\_l and 2f\_bhabhag, respectively,
and background with $\mu\mu\nu\nu$ or $\tau\tau\nu\nu$
as the final state is denoted with 4f\_zzorww\_l. First, a loose precut on $M_{\mathrm{rec}}$ is
applied as $M_{\mathrm{rec}}\in${[}100, 300{]} GeV. Then the following
cuts are applied in this order:
\begin{itemize}
\item since the invariant mass $M_{\mathrm{l^{+}l^{-}}}$ ($\mathrm{l}$
= $\mathrm{e}$ or $\mu$) of the dilepton system should be close
to the Z boson mass for the Higgsstrahlung process, a criterion is
imposed as $M_{\mathrm{l^{+}l^{-}}}\in${[}73, 120{]} GeV. The top
plot in Figure \ref{fig:-Minv} compares the $M_{\mathrm{l^{+}l^{-}}}$
of signal and major background processes.
\item for the signal, the transverse dilepton momentum $p_{\mathrm{T}}^{\mathrm{l^{+}l^{-}}}$
should peak at a certain value determined by kinematics. In contrast,
the $p_{\mathrm{T}}^{\mathrm{l^{+}l^{-}}}$ of the two-fermion background
peaks towards small values. This motivates the cut $p_{\mathrm{T}}^{\mathrm{l^{+}l^{-}}}$>
10 GeV. In addition, an upper limit on $p_{\mathrm{T}}^{\mathrm{l^{+}l^{-}}}$
is imposed to suppress background processes whose $p_{\mathrm{T}}^{\mathrm{l^{+}l^{-}}}$
extend to large values. The center plot in Figure \ref{fig:-Minv}
compares the $p_{\mathrm{T}}^{\mathrm{l^{+}l^{-}}}$ of the signal
and major background processes.
\item $\theta_{\mathrm{missing}}$, the polar angle of the missing momentum,
discriminates against events which are unbalanced in longitudinal
momentum, in particular 2-fermion events in which ISR emitted approximately
collinear with the incoming beams escapes detection in the beam pipe.
The bottom plot in Figure \ref{fig:-Minv} shows the distribution of
$\mathrm{cos\left(\theta_{missing}\right)}$ between the signal and
major background processes. A cut is made at $\mathrm{\left|cos\left(\theta_{missing}\right)\right|}<0.98$,
which cuts 2-fermion backgrounds by approximately two thirds.
\item multi-variate cut: While the $p_{\mathrm{T}}^{\mathrm{l^{+}l^{-}}}$
and $\mathrm{cos\left(\theta_{missing}\right)}$ cuts are effective
for removing 2-fermion backgrounds, the signatures of 4-fermion backgrounds
are harder to distinguish from the Higgsstrahlung signal. Nevertheless,
further rejection of residual background events is achieved by a multi-variate
(MVA) cut based on the Boosted Decision Tree (BDT) method \cite{Hocker:2007ht}
using a combination of the variables $M_{\mathrm{l^{+}l^{-}}}$, $\mathrm{cos\left(\theta_{Z}\right)}$,
$\mathrm{cos\left(\theta_{lep}\right)}$, $\mathrm{cos\left(\theta_{track,1}\right)}$
and $\mathrm{cos\left(\theta_{track,2}\right)}$. Here, $\theta_{\mathrm{Z}}$
is the polar angle of the Z boson, $\theta_{\mathrm{lep}}$ is the
angle between the leptons, and $\theta_{\mathrm{track,1,2}}$ is the
polar angle of each lepton track. The BDT response is calculated using
weights obtained from training samples consisting of simulated signal
and background events. Figure \ref{fig:likelihood-Var} shows the
distribution of the variables used for the MVA training, as well as
the BDT response for signal and background. The MVA cut is optimized
for each channel to maximize $\sigma_{\mathrm{ZH}}$ precision.
\item recoil mass cut: $\sigma_{\mathrm{ZH}}$ and $M_{\mathrm{H}}$ are
obtained by fitting the $M_{\mathrm{rec}}$ spectrum within a wide
window around the signal $M_{\mathrm{rec}}$ peak. This is designated
to be $M_{\mathrm{rec}}\in${[}110, 155{]} GeV for $\sqrt{s}$=250
GeV, {[}100, 200{]} GeV for $\sqrt{s}$=350 GeV, and {[}100, 250{]}
GeV for $\sqrt{s}$= 500 GeV.
\item visible energy cut: $E_{\mathrm{vis}}$, defined as the visible energy
excluding that from the isolated lepton pair, is required to be above
a certain value (10 GeV for $\sqrt{s}$=250 and 350 GeV and 25 GeV
for $\sqrt{s}$=500 GeV) in order to suppress one of the dominant
residual backgrounds which has $\mathrm{ll\nu\nu}$ ($\mathrm{l}$
= $\mathrm{e}$ or $\mu$) in the final state. The distributions of
$E_{\mathrm{vis}}$ are compared between signal and $\mathrm{ll\nu\nu}$
background in Figure \ref{fig:Evis}. The improvement on $\sigma_{\mathrm{ZH}}$
and $M_{\mathrm{H}}$ is significant in the case of the $\mathrm{e_{L}^{-}e_{R}^{+}}$
polarization\cite{Yan:2016trc}, where the contribution of $\mathrm{ll\nu\nu}$
background with WW intermediate states is large. Although the $E_{\mathrm{vis}}$
cut also excludes signal events in which the Higgs boson decays invisibly,
Higgs decay model independence is maintained by combining the results
obtained from this analysis with a dedicated analysis for invisible
Higgs decays \cite{JunpingTian2015:Invisible,AkimasaIshikawa2014:invisibleHad}. This is explained by
the fact that the $\mathrm{ZH}$ cross section for the SM Higgs boson
can be expressed as $\sigma_{\mathrm{ZH}}=\sigma_{\mathrm{ZH,vis}}+\sigma_{\mathrm{\mathrm{ZH,inv}is}}$,
where $\sigma_{\mathrm{ZH,vis}}$ and $\sigma_{\mathrm{ZH,invis}}$
, which are the cross sections of the visible and invisible decay
events, respectively, can both be measured individually and model
independently.
\end{itemize}
For the case of $\sqrt{s}$=250 GeV, Tables \ref{tab:cut eff} and
\ref{tab:cut effZee} show the number of remaining signal and background,
signal efficiency and significance after each cut. Similar outcomes
are obtained for $\sqrt{s}$=350 and 500 GeV since similar data selection
methods are used. For the case of $\sqrt{s}$=250 GeV, Figure \ref{fig:stk}
shows distributions of the $M_{\mathrm{rec}}$ of the signal and major
residual background processes, which are 4f\_sl and 2f\_l defined
in Section \ref{sub:Signal-and-Background}. Figures \ref{fig:rec250}
- \ref{fig:rec500} show the reconstructed $M_{\mathrm{rec}}$ spectra
of the events remaining in a wide region around the signal $M_{\mathrm{rec}}$
peak for all three center-of-mass energies. Only the plots for $\mathrm{e_{L}^{-}e_{R}^{+}}$
are shown for $\sqrt{s}$=350 and 500 GeV for the sake of brevity.
The following can be observed:
\begin{itemize}
\item A sharper signal peak and a better signal-to-background ratio can
be achieved at a smaller center-of-mass energy. This is explained
by (a) the Higgsstrahlung cross section maximizes near $\sqrt{s}$=250
GeV, then decreases with energy, (b) the detector momentum resolution
degrades linearly with momentum, and (c) the larger beamsstrahlung
effect at higher center-of-mass energies enhances the tail of the
$M_{\mathrm{rec}}$ spectra for both signal and background processes.
\item The $\mathrm{\mu^{+}\mu^{-}H}$ channel has a sharper signal
peak hence better mass resolution than the $\mathrm{e^{+}e^{-}H}$
channel which suffers from bremsstrahlung.
\item $\mathrm{e_{L}^{-}e_{R}^{+}}$ benefits
from larger signal cross section, whereas the background level is
lower for $\mathrm{e_{R}^{-}e_{L}^{+}}$
since the background events from WW processes are significantly suppressed.
\end{itemize}
These traits account for the precision of $\sigma_{\mathrm{ZH}}$
and $M_{\mathrm{H}}$ evaluated in Section \ref{sub:Discussion-of-the}.

\begin{table*}
\centering
\caption{The number of events left after each cut for the $\mathrm{\mu^{+}\mu^{-}H}$
channel and $\mathrm{e_{L}^{-}e_{R}^{+}}$ at $\sqrt{s}$=250 GeV.
Also given are the efficiency and signal significance (defined as
$\frac{N_{S}}{\sqrt{N_{S}+N_{B}}}$ ,where $N_{S(B)}$ is the number
of signal (background)) for the Higgsstrahlung signal. Precut represents
the loose cut $M_{\mathrm{rec}}\in${[}100, 300{]} GeV.}
\label{tab:cut eff}

\begin{tabular}{|c|c|c|c|c|c|c|c|}
\hline
$\int\mathcal{L}dt$  & $\mathrm{\mu^{+}\mu^{-}H}$  & signal & signal &  &  &  & total\tabularnewline
= 250 $\mathrm{fb^{-1}}$ & $\mathrm{e_{L}^{-}e_{R}^{+}}$ & efficiency & significance & 2f\_l & 4f\_l & 4f\_sl & background\tabularnewline
\hline
\hline
no cut & 2603 & 100\% & 0.42 & 9.54$\times10^{6}$ & 3.15$\times10^{6}$ & 4.98$\times10^{6}$ & 1.98$\times10^{7}$\tabularnewline
\hline
Lepton ID+Precut & 2439 & 93.70\% & 7.46 & 61675 & 34451 & 8218 & 104344\tabularnewline
\hline
$M_{\mathrm{l^{+}l^{-}}}\in${[}73, 120{]} GeV & 2382 & 91.51\% & 8.09 & 54352 & 22543 & 7446 & 84341\tabularnewline
\hline
$p_{\mathrm{T}}^{\mathrm{l^{+}l^{-}}}\in${[}10, 70{]} GeV & 2335 & 89.70\% & 11.17 & 15429 & 19648 & 6245 & 41322\tabularnewline
\hline
$\mathrm{\left|\cos\theta_{\text{missing}}\right|}$< 0.98 & 2335 & 89.70\% & 12.71 & 5594 & 19539 & 6245 & 31378\tabularnewline
\hline
BDT > - 0.25 & 2310 & 88.74\% & 15.03 & 4195 & 12530 & 4586 & 21311\tabularnewline
\hline
$M_{\mathrm{rec}}\in${[}110, 155{]} GeV & 2296 & 88.21\% & 16.37 & 3522 & 10423 & 3433 & 17378\tabularnewline
\hline
$E_{\mathrm{vis}}$ > 10 GeV & 2293 & 88.09\% & 20.94 & 3261 & 2999 & 3433 & 9694\tabularnewline
\hline
\end{tabular}
\end{table*}

\begin{table*}
\centering
\caption{The number of events left after each cut for the $\mathrm{e{}^{+}e{}^{-}H}$
channel and $\mathrm{e_{L}^{-}e_{R}^{+}}$ at $\sqrt{s}$=250 GeV.
Also given are the efficiency and signal significance for the Higgsstrahlung
signal.}
\label{tab:cut effZee}

\begin{tabular}{|c|c|c|c|c|c|c|c|}
\hline
$\int\mathcal{L}dt$  & $\mathrm{e^{+}e^{-}H}$  & signal & signal &  &  &  & total\tabularnewline
= 250 $\mathrm{fb^{-1}}$ & $\mathrm{e_{L}^{-}e_{R}^{+}}$ & efficiency & significance & 2f\_l & 4f\_l & 4f\_sl & background\tabularnewline
\hline
\hline
no cut & 2729 & 100\% & 0.44 & 9.54$\times10^{6}$ & 3.15$\times10^{6}$ & 4.98$\times10^{6}$ & 1.98$\times10^{7}$\tabularnewline
\hline
Lepton ID+Precut & 2422 & 86.99\% & 4.83 & 181196 & 51406 & 16093 & 248929\tabularnewline
\hline
$M_{\mathrm{l^{+}l^{-}}}\in${[}73, 120{]} GeV & 2351 & 84.50\% & 6.24 & 99934 & 28612 & 10876 & 139581\tabularnewline
\hline
$p_{\mathrm{T}}^{\mathrm{l^{+}l^{-}}}\in${[}10, 70{]} GeV & 2300 & 84.28\% & 6.78 & 79066 & 24425 & 9289 & 112933\tabularnewline
\hline
$\mathrm{\left|\cos\theta_{\text{missing}}\right|}$< 0.98 & 2300 & 84.24\% & 8.63 & 35299 & 23931 & 9261 & 66844\tabularnewline
\hline
BDT > 0.019 & 1860 & 68.15\% & 14.95 & 5000 & 5370 & 3229 & 13624\tabularnewline
\hline
$M_{\mathrm{rec}}\in${[}110, 155{]} GeV & 1853 & 67.90\% & 15.90 & 4390 & 4791 & 2522 & 11728\tabularnewline
\hline
$E_{\mathrm{vis}}$ > 10 GeV & 1850 & 67.79\% & 18.58 & 4326 & 1190 & 2522 & 8062\tabularnewline
\hline
\end{tabular}
\end{table*}

\section{Extraction of Higgs Mass and Higgs Production Cross Section}
\label{sec:Extraction}
This section presents the methods to extract the Higgs boson mass
($M_{\mathrm{H}}$) and the cross section ($\sigma_{\mathrm{ZH}}$)
and discusses the results.

\subsection{Fitting method}
\label{sub:Method-of-Fitting}

\begin{figure}
\centering
\includegraphics[width=\linewidth]{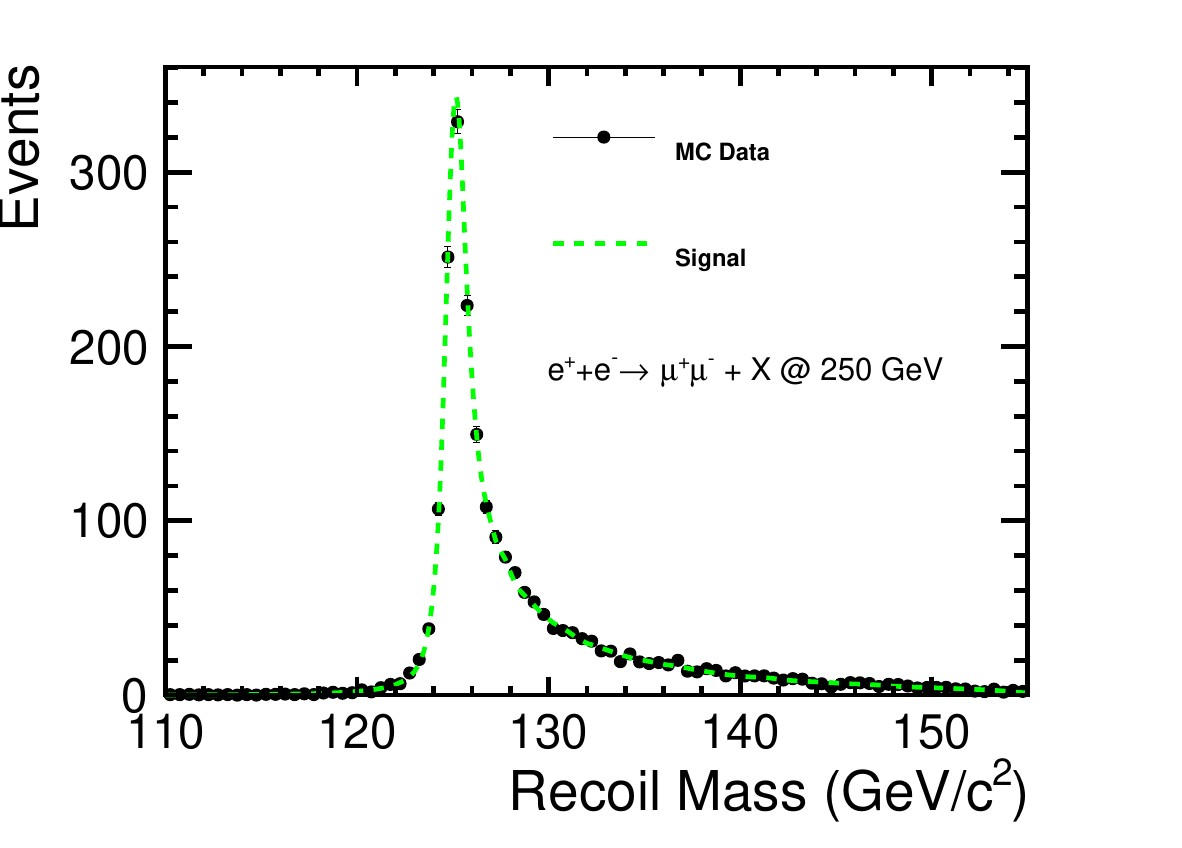}
\includegraphics[width=\linewidth]{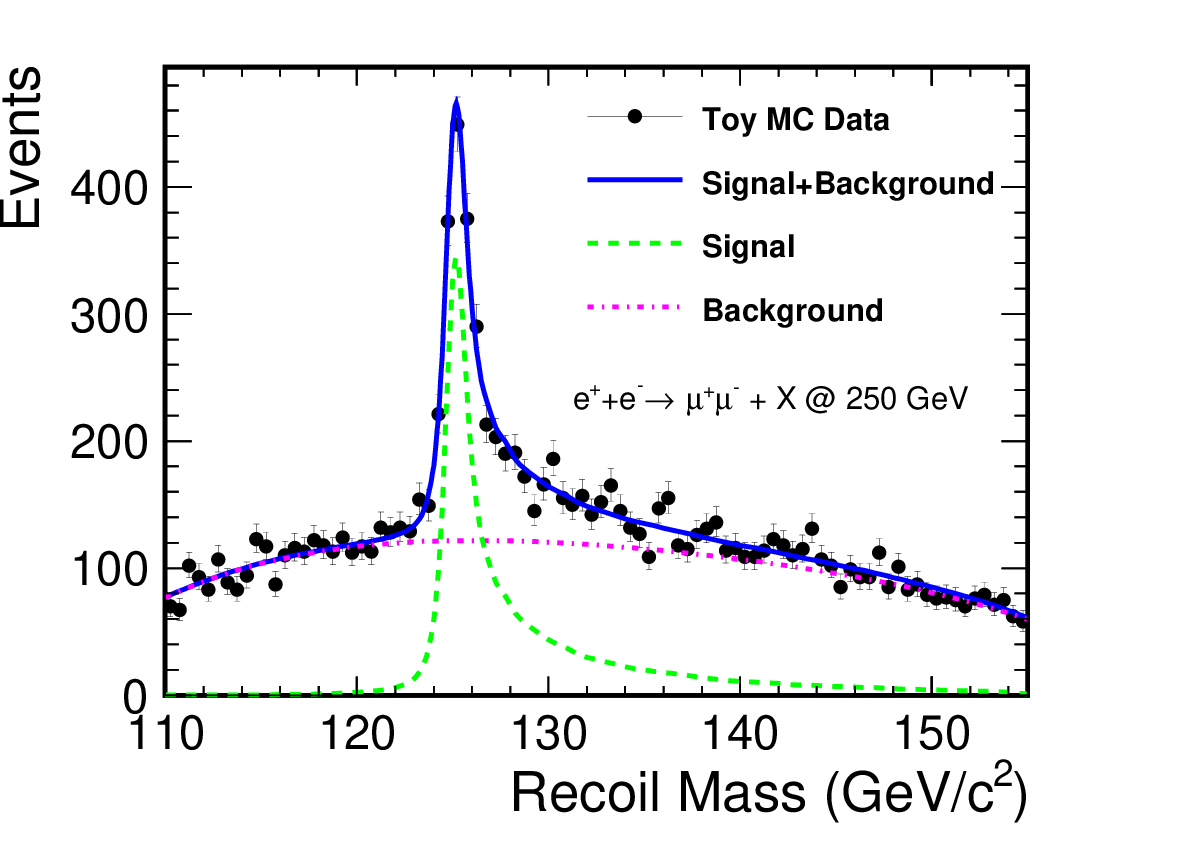}
\includegraphics[width=\linewidth]{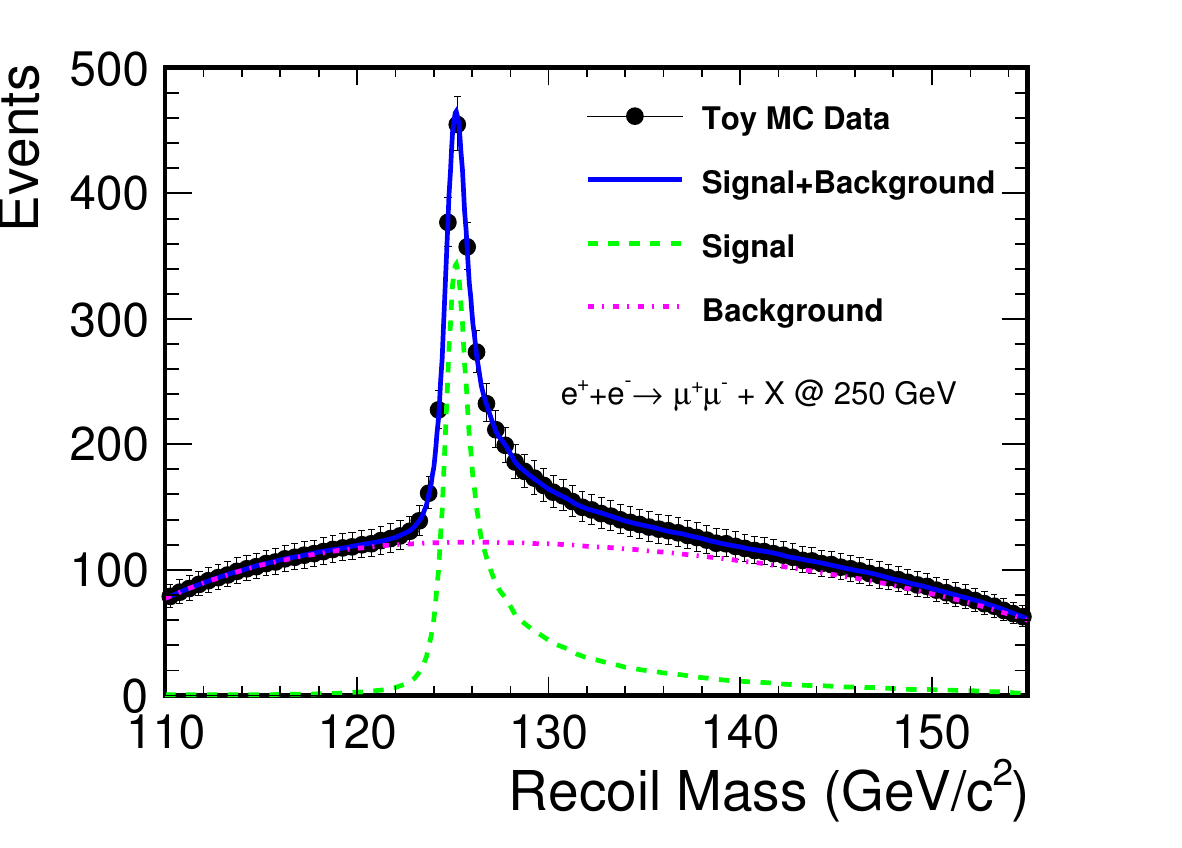}

\caption{For the case of the $\mathrm{\mu^{+}\mu^{-}H}$ channel
and $\mathrm{e_{L}^{-}e_{R}^{+}}$ at
$\sqrt{s}$ = 250 GeV, in the region 110-155 GeV: (top) The $M_{\mathrm{rec}}$
spectra of the signal MC events used in analysis plotted together
with the kernel function. (center) The $M_{\mathrm{rec}}$ spectrum
of toy MC events corresponding to the top plot. (bottom) Toy
MC events used for extracting $\sigma_{\mathrm{ZH}}$ and $M_{\mathrm{H}}$
and their statistical uncertainties, which are generated using the
function which fitted the top plot as input. The legend is the same as in Figure \ref{fig:rec250}.}
\label{fig:rec250-1}
\end{figure}

After applying the selection introduced in the previous section, the
remaining $M_{\mathrm{rec}}$ spectrum is a superposition of signal
and residual background events. The observables of interest, $\sigma_{\mathrm{ZH}}$
and $M_{\mathrm{H}}$, are extracted by fitting the MC data using
a multi-component function in a wide region surrounding the signal
peak. These are shown in Figures \ref{fig:rec250} - \ref{fig:rec500}.
The signal spectrum is modeled in a non-parametric way using a Gaussian
kernel estimation method \cite{Cranmer:2000du}. Figure \ref{fig:rec250-1}
(top) shows the $M_{\mathrm{rec}}$ spectrum of the signal MC data
plotted together with the kernel function ($F_{\mathrm{S}}$). The
kernel function shape does not change with variations in the Higgs
boson mass within a range of about 1-2 GeV, hence $M_{\mathrm{H}}$
can be obtained as a free parameter by allowing the kernel function
to shift in the fitting process. The background spectrum is approximated
by either a third or fourth order Chebyshev polynomial ($F_{B}$),
depending on the shape of the distribution for each channel. The MC
data is fitted as a sum of the kernel function and the Chebyshev polynomial
by $F_{\mathrm{tot}}\left(x,\:M_{\mathrm{H}}\right)=N_{\mathrm{S}}\cdot F_{\mathrm{S}}\left(x,\:M_{\mathrm{H}}\right)+N_{\mathrm{B}}\cdot F_{\mathrm{B}}\left(x,c_{i}\right)$.
Here, $N_{\mathrm{S}}$ is the signal yield and $M_{\mathrm{H}}$
is the mass parameter ($M_{\mathrm{H}}$=125 GeV for the signal sample
used to obtain the kernel function); $N_{\mathrm{B}}$ is the background
yield, and $c_{i}$ ($i$=0, 1,..., 3 or 4 corresponding to 3rd or
4th order polynomial) are the coefficients of of $F_{B}$, which are
obtained from fitting the MC background only data.

The uncertainties of $\sigma_{\mathrm{ZH}}$ and $M_{\mathrm{H}}$
are evaluated using a toy MC procedure. The toy MC events (bottom
right plot in Figure \ref{fig:rec250-1}) are generated from $F_{\mathrm{tot}}$
with $M_{\mathrm{H}}$=125 GeV and $N_{\mathrm{S}}$ as input, then
fitted by $F_{\mathrm{tot}}$ with $N_{\mathrm{S}}$ and $M_{\mathrm{H}}$
floated and the background shape $F_{\mathrm{B}}$ and background
normalization $N_{\mathrm{B}}$ fixed. \footnote{Here it is assumed that the background yield can be estimated with
a reasonable precision of a few percent. If the background yield were
floated, the $\sigma_{\mathrm{ZH}}$ uncertainty would increase by
10-20\%.} The information obtained from fitting are $N_{S}$, $M_{\mathrm{H}}$
, and their statistical uncertainties ($\Delta N_{S}$ and $\Delta M_{\mathrm{H}}$).
$N_{S}$ can be translated to $\sigma_{\mathrm{ZH}}$ through

\begin{equation}
\sigma_{\mathrm{ZH}}=\frac{N_{S}}{BR\mathrm{\left(\mathrm{Z\rightarrow\mathrm{l^{+}l^{-}}}\right)\varepsilon}_{S}L}\:,\label{eq:xsec}
\end{equation}
where $\varepsilon_{\mathrm{S}}$ is the efficiency of signal event
selection, $BR\mathrm{\left(\mathrm{Z\rightarrow\mathrm{l^{+}l^{-}}}\right)}$
the branching ratio of the Z boson decaying to a pair of leptons of
type $\mathrm{l}$, and $L$ the integrated luminosity. Therefore
the relative statistical uncertainty $\Delta N_{S}/N_{S}$ is equal
to $\Delta\sigma_{\mathrm{ZH}}/\sigma_{\mathrm{ZH}}$. The shift
in the fitted value of $M_{\mathrm{H}}$ is negligible with respect
to its statistical uncertainties.

\subsection{Discussion of the results}
\label{sub:Discussion-of-the}

\subsubsection{Precision evaluation based on nominal integrated luminosities}
\label{sub:Precision-evaluation-based}

Table \ref{Resukt250=0000231-2} shows the expected precisions of
$\sigma_{\mathrm{ZH}}$ and $M_{\mathrm{H}}$ assuming the integrated
luminosities of 250 $\mathrm{fb^{-1}}$, 333 $\mathrm{fb^{-1}}$,
and 500 $\mathrm{fb^{-1}}$ for $\sqrt{s}$ = 250, 350, and 500 GeV,
respectively, for each beam polarization. In order to maintain the
model independence of the $\sigma_{\mathrm{ZH}}$ measurement, the
results in Table \ref{Resukt250=0000231-2} are combined with those
from invisible Higgs decay analyses;
Table \ref{ResultConvo} shows the combined results. \footnote{The relative uncertainties of the cross section of the invisible Higgs
decay events are assumed to be half of the upper limit on invisible
decay branching ratios given in \cite{JunpingTian2015:Invisible}. }

\begin{table}
\centering
\caption{The statistical uncertainties on $\sigma_{\mathrm{ZH}}$ and $\Delta M_{\mathrm{H}}$,
assuming for each beam polarization a total integrated luminosity
of 250 $\mathrm{fb^{-1}}$, 333 $\mathrm{fb^{-1}}$, and 500 $\mathrm{fb^{-1}}$
for $\sqrt{s}$ = 250, 350, and 500 GeV, respectively. The results
are given in the form of separate and combined results of the $\mathrm{\mu^{+}\mu^{-}X}$
and $\mathrm{e^{+}e^{-}X}$ channels. \label{Resukt250=0000231-2}}
\begin{tabular}{|c|c|c|c|c|}
\hline
\multicolumn{1}{|c}{} & $\sqrt{s}$  & 250 GeV & 350 GeV & 500 GeV\tabularnewline
\hline
\multicolumn{1}{|c}{} &  & $\Delta\sigma_{\mathrm{ZH}}/\sigma_{\mathrm{ZH}}$ & $\Delta\sigma_{\mathrm{ZH}}/\sigma_{\mathrm{ZH}}$ & $\Delta\sigma_{\mathrm{ZH}}/\sigma_{\mathrm{ZH}}$\tabularnewline
\hline
\hline
$\mathrm{e_{L}^{-}e_{R}^{+}}$ & $\mathrm{\mu^{+}\mu^{-}H}$  & 3.2\% & 3.9\% & 6.9\%\tabularnewline
\cline{2-5}
 & $\mathrm{e^{+}e^{-}H}$ & 4.0\% & 5.3\% & 7.2\%\tabularnewline
\cline{2-5}
 & combined & 2.5\% & 3.1\% & 5.0\%\tabularnewline
\hline
$\mathrm{e_{R}^{-}e_{L}^{+}}$ & $\mathrm{\mu^{+}\mu^{-}H}$  & 3.6\% & 4.5\% & 8.1\%\tabularnewline
\cline{2-5}
 & $\mathrm{e^{+}e^{-}H}$ & 4.7\% & 6.1\% & 7.5\%\tabularnewline
\cline{2-5}
 & combined & 2.9\% & 3.6\% & 5.5\%\tabularnewline
\hline
\end{tabular}

\begin{tabular}{|c|c|c|c|c|}
\hline
\multicolumn{1}{|c}{} & $\sqrt{s}$  & 250 GeV & 350 GeV & 500 GeV\tabularnewline
\hline
\multicolumn{1}{|c}{} &  & $\Delta M_{\mathrm{H}}$ (MeV) & $\Delta M_{\mathrm{H}}$ (MeV) & $\Delta M_{\mathrm{H}}$ (MeV)\tabularnewline
\hline
\hline
$\mathrm{e_{L}^{-}e_{R}^{+}}$ & $\mathrm{\mu^{+}\mu^{-}H}$  & 39 & 103 & 592\tabularnewline
\cline{2-5}
 & $\mathrm{e^{+}e^{-}H}$ & 121 & 450 & 1160\tabularnewline
\cline{2-5}
 & combined & 37 & 100 & 527\tabularnewline
\hline
$\mathrm{e_{R}^{-}e_{L}^{+}}$ & $\mathrm{\mu^{+}\mu^{-}H}$  & 43 & 120 & 660\tabularnewline
\cline{2-5}
 & $\mathrm{e^{+}e^{-}H}$ & 149 & 502 & 1190\tabularnewline
\cline{2-5}
 & combined & 41 & 117 & 577\tabularnewline
\hline
\end{tabular}
\end{table}

\begin{table}
\centering
\caption{The model independent statistical uncertainties on $\sigma_{\mathrm{ZH}}$
obtained by combining the results of $\Delta\sigma_{\mathrm{ZH}}/\sigma_{\mathrm{ZH}}$
in Table \ref{Resukt250=0000231-2} with those of the invisible Higgs
decay analysis, assuming for each beam polarization
a total integrated luminosity of 250 $\mathrm{fb^{-1}}$, 333 $\mathrm{fb^{-1}}$,
and 500 $\mathrm{fb^{-1}}$ for $\sqrt{s}$ = 250, 350, and 500 GeV,
respectively.}
\label{ResultConvo}

\begin{tabular}{|c|c|c|c|c|}
\hline
Pol. & $\sqrt{s}$ & 250 GeV & 350 GeV & 500 GeV\tabularnewline
\hline
\hline
$\mathrm{e_{L}^{-}e_{R}^{+}}$ & $\Delta\sigma_{\mathrm{ZH}}/\sigma_{\mathrm{ZH}}$ & 2.5\% & 3.2\% & 5.1\%\tabularnewline
\hline
$\mathrm{e_{R}^{-}e_{L}^{+}}$ & $\Delta\sigma_{\mathrm{ZH}}/\sigma_{\mathrm{ZH}}$ & 2.9\% & 3.6\% & 5.6\%\tabularnewline
\hline
\end{tabular}
\end{table}

\subsubsection{Impact of center-of-mass energy and beam polarization}

Table \ref{ResultCompare} compares the precisions of higher $\sqrt{s}$
= 350 and 500 GeV with respect to $\sqrt{s}$ = 250 GeV, as well as
the precisions of beam polarization $\mathrm{e_{R}^{-}e_{L}^{+}}$
to that of $\mathrm{e_{L}^{-}e_{R}^{+}}$. The same integrated luminosities
as those mentioned in Section \ref{sub:Precision-evaluation-based}
are assumed. The following can be observed:
\begin{itemize}
\item Compared to $\sqrt{s}$=250 GeV, the precision of $\sigma_{\mathrm{ZH}}$
at $\sqrt{s}$=350 GeV is worse by about a factor of 1.3, while $\Delta M_{\mathrm{H}}$
is worse by a factor of about 2.7.
\item Compared to $\sqrt{s}$=250 GeV, the precision of $\sigma_{\mathrm{ZH}}$
at $\sqrt{s}$=500 GeV is worse by a factor of about 2.1, while $\Delta M_{\mathrm{H}}$
is worse by a factor of about 14.
\item In general, the precision of $\mathrm{e_{L}^{-}e_{R}^{+}}$ is worse
by a factor of 1.1 - 1.2 with respect to that of $\mathrm{e_{L}^{-}e_{R}^{+}}$.
\end{itemize}

\begin{table}
\centering
\caption{The ratio of the uncertainties of $\sigma_{\mathrm{ZH}}$ and $\Delta M_{\mathrm{H}}$
for $\sqrt{s}$=350 and 500 GeV with respect to $\sqrt{s}$=250 GeV
(top), as well for $\mathrm{e_{R}^{-}e_{L}^{+}}$ with respect to
$\mathrm{e_{L}^{-}e_{R}^{+}}$(bottom). These are based on the results
given in Tables \ref{Resukt250=0000231-2} and \ref{ResultConvo},
which assume for each beam polarization a total luminosity of 250
$\mathrm{fb^{-1}}$, 333 $\mathrm{fb^{-1}}$, and 500 $\mathrm{fb^{-1}}$
for $\sqrt{s}$ = 250, 350, and 500 GeV, respectively.\label{ResultCompare}}

\begin{tabular}{|c|c|c|c|c|}
\hline
 \multicolumn{2}{|c|}{$\sqrt{s}$} & 250 GeV & 350 GeV & 500 GeV\tabularnewline
\hline
\multicolumn{2}{|c|}{$\int\mathcal{L}dt$} & 250 $\mathrm{fb^{-1}}$ & 333 $\mathrm{fb^{-1}}$ & 500 $\mathrm{fb^{-1}}$\tabularnewline
\hline
\hline
wrt. & $\Delta\sigma_{\mathrm{ZH}}/\sigma_{\mathrm{ZH}}$ & 1 & 1.3 x & 2.1 x\tabularnewline
\cline{2-5}
  $\sqrt{s}$=250 GeV & $\Delta M_{\mathrm{H}}$ & 1 & 2.7 x & 14 x\tabularnewline
\hline
\hline
$\mathrm{e_{R}^{-}e_{L}^{+}}$  wrt.  & $\Delta\sigma_{\mathrm{ZH}}/\sigma_{\mathrm{ZH}}$ & 1.1 x & 1.2 x & 1.1 x\tabularnewline
\cline{2-5}
  $\mathrm{e_{L}^{-}e_{R}^{+}}$ & $\Delta M_{\mathrm{H}}$ & 1.1 x & 1.2 x & 1.1 x\tabularnewline
\hline
\end{tabular}

\end{table}

\subsubsection{Scaled to the H20 run scenario}

Table \ref{ResultConvoH20} shows the uncertainties of $\sigma_{\mathrm{ZH}}$
(from Table \ref{ResultConvo}) and $M_{\mathrm{H}}$ scaled to the
full H20 run scenario\cite{Fujii:2015jha,Barklow:2015tja}. A total of 2000 $\mathrm{fb^{-1}}$
, 200 $\mathrm{fb^{-1}}$, and 4000 $\mathrm{fb^{-1}}$ are accumulated
at $\sqrt{s}$ = 250, 350, and 500 GeV, respectively, out of which
67.5\% (22.5\%) of the running time are dedicated to $\mathrm{e_{L}^{-}e_{R}^{+}}$
($\mathrm{e_{R}^{-}e_{L}^{+}}$) at $\sqrt{s}$ = 250 and 350 GeV,
while 40\% of the running time is dedicated to each of $\mathrm{e_{L}^{-}e_{R}^{+}}$
and $\mathrm{e_{L}^{-}e_{R}^{+}}$ at $\sqrt{s}$ = 500 GeV.

From each measurement of $\sigma_{\mathrm{ZH}}$, the $\mathrm{HZZ}$
coupling ($g_{\mathrm{HZZ}}$) can be obtained based on $\sigma_{\mathrm{ZH}}\propto g_{\mathrm{HZZ}}^{2}$,
which results in $\Delta g_{\mathrm{HZZ}}/g_{\mathrm{HZZ}}=\frac{1}{2}\cdot\Delta\sigma_{\mathrm{ZH}}/\sigma_{\mathrm{ZH}}$.
Table \ref{ResultConvoH20} gives the combined errors of $\Delta g_{\mathrm{HZZ}}/g_{\mathrm{HZZ}}$
and $\Delta M_{\mathrm{H}}$. It can be seen that from the leptonic
recoil measurements alone, a precision of 0.4\% and 14 MeV can be
achieved for $\Delta g_{\mathrm{HZZ}}/g_{\mathrm{HZZ}}$ and $M_{\mathrm{H}}$,
respectively by the end of the 20 year run, with the dominant contribution
from $\sqrt{s}$ = 250 GeV.

\begin{table*}
\centering
\caption{The uncertainties of $\sigma_{\mathrm{ZH}}$ and $M_{\mathrm{H}}$
scaled to the full H20 run scenario, as well
as the combined errors of $\Delta g_{\mathrm{HZZ}}/g_{\mathrm{HZZ}}$
and $\Delta M_{\mathrm{H}}$.}
\label{ResultConvoH20}

\begin{tabular}{|c|c|c|c|c|c|c|}
\hline
$\sqrt{s}$  & \multicolumn{1}{c}{250 GeV} &  & \multicolumn{1}{c}{350 GeV} &  & \multicolumn{1}{c}{500 GeV} & \tabularnewline
\hline
 & $\int\mathcal{L}dt$  & $\Delta\sigma_{\mathrm{ZH}}/\sigma_{\mathrm{ZH}}$ & $\int\mathcal{L}dt$  & $\Delta\sigma_{\mathrm{ZH}}/\sigma_{\mathrm{ZH}}$ & $\int\mathcal{L}dt$  & $\Delta\sigma_{\mathrm{ZH}}/\sigma_{\mathrm{ZH}}$\tabularnewline
\hline
\hline
$\mathrm{e_{L}^{-}e_{R}^{+}}$ & 1350 $\mathrm{fb^{-1}}$ & 1.1\% & 115 $\mathrm{fb^{-1}}$ & 5.0\% & 1600 $\mathrm{fb^{-1}}$ & 2.9\%\tabularnewline
\hline
$\mathrm{e_{R}^{-}e_{L}^{+}}$ & 450 $\mathrm{fb^{-1}}$ & 2.2\% & 45 $\mathrm{fb^{-1}}$ & 9.8\% & 1600 $\mathrm{fb^{-1}}$ & 3.1\%\tabularnewline
\hline
\multicolumn{1}{c|}{} & \multicolumn{6}{c|}{H20 combined: $\Delta g_{\mathrm{ZZH}}/g_{\mathrm{ZZH}}$ = 0.4\%} \tabularnewline
\cline{2-7}
\end{tabular}

\begin{tabular}{|c|c|c|c|c|c|c|}
\hline
$\sqrt{s}$  & \multicolumn{1}{c}{250 GeV} &  & \multicolumn{1}{c}{350 GeV} &  & \multicolumn{1}{c}{500 GeV} & \tabularnewline
\hline
 & $\int\mathcal{L}dt$  & $\Delta M_{\mathrm{H}}$ (MeV) & $\int\mathcal{L}dt$  & $\Delta M_{\mathrm{H}}$ (MeV) & $\int\mathcal{L}dt$  & $\Delta M_{\mathrm{H}}$ (MeV)\tabularnewline
\hline
\hline
$\mathrm{e_{L}^{-}e_{R}^{+}}$ & 1350 $\mathrm{fb^{-1}}$ & 16 & 115 $\mathrm{fb^{-1}}$ & 157 & 1600 $\mathrm{fb^{-1}}$ & 295\tabularnewline
\hline
$\mathrm{e_{R}^{-}e_{L}^{+}}$ & 450 $\mathrm{fb^{-1}}$ & 31 & 45 $\mathrm{fb^{-1}}$ & 318 & 1600 $\mathrm{fb^{-1}}$ & 323\tabularnewline
\hline
total & 1800 $\mathrm{fb^{-1}}$ & 14 & 160 $\mathrm{fb^{-1}}$ & 141 & 3200 $\mathrm{fb^{-1}}$ & 218\tabularnewline
\hline
\multicolumn{1}{c|}{} & \multicolumn{6}{c|}{H20 combined: $\Delta M_{\mathrm{H}}$ = 14 MeV} \tabularnewline
\cline{2-7}
\end{tabular}
\end{table*}

\section{Demonstration of Higgs Decay Mode Independence}
\label{sec:ModeBias}

In the recoil method, $\sigma_{\mathrm{ZH}}$ is measured without
any explicit assumption regarding Higgs decay modes. This section
demonstrates that the $\sigma_{\mathrm{ZH}}$ measured using the methods
described in previous sections does not depend on the underlying model
which determines the Higgs decay modes and their branching ratios.
More details on this study are given in\cite{Yan:2016trc}. The
key question here is whether the $\sigma_{\mathrm{ZH}}$ extracted
in Equation \ref{eq:xsec} using the measured number of signal events
($N_{\mathrm{S}}$) and the signal selection efficiency ($\varepsilon_{\mathrm{S}}$)
from the Monte Carlo samples would be biased when the Higgs boson
decays differently from that assumed in the samples.

First we introduce the general strategies towards a model independent
$\sigma_{\mathrm{ZH}}$ measurement. The direct observable $N_{\mathrm{S}}$
can be parameterised as

\begin{equation}
N_{\mathrm{S}}=\underset{i}{\varSigma}N_{i}=\underset{i}{\varSigma}\sigma_{\mathrm{ZH}}R_{l}LB_{i}\varepsilon_{i}\,,\label{eq:BR1}
\end{equation}
where the summation goes through all Higgs decay modes. $N_{\mathrm{i}}$,
$B_{\mathrm{i}}$, and $\varepsilon_{\mathrm{i}}$ are the the number
of signal events, branching ratio and selection efficiency of Higgs
decay mode $i$, respectively. $L$ is the integrated luminosity,
and $R_{l}$ is the branching ratio of $\mathrm{Z\rightarrow l^{+}l^{-}}$.
If the signal efficiency equals to the same $\varepsilon$ for all
decay modes, Equation \ref{eq:BR1} becomes

\begin{equation}
N_{\mathrm{S}}=\sigma_{\mathrm{ZH}}R_{l}L\varepsilon\underset{i}{\varSigma}B_{i}\,.\label{eq:BR1-1}
\end{equation}
Since $\underset{i}{\varSigma}B_{i}=1$ stands in any case, $\sigma_{\mathrm{ZH}}$
can be extracted without assumptions on decay modes or branching ratios
as

\begin{equation}
\sigma_{\mathrm{ZH}}=\frac{N_{\mathrm{S}}}{R_{l}L\varepsilon}\,,\label{eq:BR2-2}
\end{equation}
This is the ideal case which guarantees model independence. On the
other hand, if there exist discrepancies between the signal efficiencies
of each mode, $\sigma_{\mathrm{ZH}}$ has to be extracted as

\begin{equation}
\sigma_{\mathrm{ZH}}=\frac{N_{\mathrm{S}}}{R_{l}L\underset{i}{\varSigma}B_{i}\varepsilon_{i}}\equiv\frac{N_{\mathrm{S}}}{R_{l}L\overline{\varepsilon}}\,,\label{eq:BR2}
\end{equation}
where $\overline{\varepsilon}=\underset{i}{\varSigma}B_{i}\varepsilon_{i}$
is the expected efficiency for all decay modes. In this case, the
bias on $\sigma_{\mathrm{ZH}}$ depends on the determination of $\overline{\varepsilon}$.
This is discussed as follows in terms of three possible scenarios
of our knowledge of Higgs decay at the time of $\sigma_{\mathrm{ZH}}$
measurement.
\begin{itemize}
\item scenario A: all Higgs decay modes and the corresponding $B_{i}$ for
each mode are known. In this rather unlikely case, $\overline{\varepsilon}$
can be determined simply by summing up over all modes, leaving no
question of model independence.
\item scenario B: $B_{i}$ is completely unknown for every mode. We would
examine the discrepancy in $\epsilon_{i}$ by investigating as many
modes as possible, and retrieve the maximum and minimum of $\epsilon_{i}$
as $\varepsilon_{\mathrm{min}}\leq\epsilon_{i}\leq\varepsilon_{\mathrm{max}}$,
from which $\overline{\varepsilon}$ can be constrained as $\varepsilon_{\mathrm{min}}\underset{i}{\Sigma}B_{i}\leq\overline{\varepsilon}\leq\varepsilon_{\mathrm{max}}\underset{i}{\Sigma}B_{i}$.
Given that $\underset{i}{\varSigma}B_{i}=1$, this can be rewritten
as $\varepsilon_{\mathrm{min}}\leq\overline{\varepsilon}\leq\varepsilon_{\mathrm{max}}$.
Then from Equation \ref{eq:BR2}, $\sigma_{\mathrm{ZH}}$ can be constrained
as
\end{itemize}
\begin{equation}
\frac{N_{\mathrm{S}}}{R_{l}L\varepsilon_{\mathrm{max}}}\leq\sigma_{\mathrm{ZH}}\leq\frac{N_{\mathrm{S}}}{R_{l}L\varepsilon_{\mathrm{min}}}\,,\label{eq:BR2-1}
\end{equation}

which indicates that the possible relative bias on $\sigma_{\mathrm{ZH}}$
can be estimated as $\frac{\varepsilon_{\mathrm{\mathrm{max}}}-\varepsilon_{\mathrm{min}}}{\varepsilon_{\mathrm{\mathrm{max}}}+\varepsilon_{\mathrm{min}}}$.
This scenario is based on a considerably conservative assumption.
\begin{itemize}
\item scenario C: $B_{i}$ is known for some of the decay modes. Here, it
is assumed that the decay modes $i$ = 1 to $n$ with a total branching
ratio of $B_{0}=\stackrel[i=1]{n}{\Sigma}B_{i}$ are known, and that
the modes from $i$ = $n+1$ with a total branching ratio of $B_{x}=\underset{i=n+1}{\Sigma}B_{i}$
are unknown. In this case, we would know the efficiency of the known
modes as $\varepsilon_{0}=\frac{\stackrel[i=1]{n}{\Sigma}B_{i}\varepsilon_{i}}{B_{0}}$.
Meanwhile the efficiency for each unknown mode can be expressed as
$\varepsilon_{i}=\varepsilon_{0}+\delta\varepsilon_{i}$, where $\delta\varepsilon_{i}$
is the deviation in efficiency for each unknown mode $i$ from $\varepsilon_{0}$.
We can then write $\overline{\varepsilon}$ as
\end{itemize}
\begin{align}
\overline{\varepsilon} &= \stackrel[i=1]{n}{\Sigma}B_{i}\varepsilon_{i}+\underset{i=n+1}{\Sigma}B_{i}\varepsilon_{i}=B_{0}\varepsilon_{0}+B_{x}\varepsilon_{0}+\underset{i=n+1}{\Sigma}B_{i}\delta\varepsilon_{i} \nonumber\\ &=\varepsilon_{0}+\underset{i=n+1}{\Sigma}B_{i}\delta\varepsilon_{i}\,.
\end{align}

The relative bias for $\overline{\varepsilon}$ and hence for $\sigma_{\mathrm{ZH}}$
is a combination of the contribution from the unknown modes and the
known modes. The contribution from the unknown modes is derived as

\begin{equation}
\frac{\Delta\sigma_{\mathrm{ZH}}}{\sigma_{\mathrm{ZH}}}=\frac{\Delta\overline{\varepsilon}}{\overline{\epsilon}}<\underset{i=n+1}{\Sigma}B_{i}\frac{\delta\varepsilon_{\mathrm{max}}}{\varepsilon_{0}}=B_{x}\frac{\delta\varepsilon_{\mathrm{max}}}{\varepsilon_{0}}\,,\label{eq:BR3}
\end{equation}
where $\delta\varepsilon_{\mathrm{max}}$ is the maximum of $\left|\delta\varepsilon_{i}\right|$
for the unknown modes. As for the known modes, because $\overline{\varepsilon}=\stackrel[i=1]{n}{\Sigma}B_{i}\varepsilon_{i}=\stackrel[i=1]{n}{\Sigma}B_{i}\left(\varepsilon_{0}+\delta\varepsilon_{i}\right)$,
where $\delta\varepsilon_{i}\equiv\varepsilon_{i}-\varepsilon_{0}$
is the deviation in efficiency for each known mode, the uncertainty
due to a fluctuation in their branching ratios ($\Delta B_{i}$) can
be expressed as $\Delta\overline{\varepsilon}=\stackrel[i=1]{n}{\Sigma}\Delta B_{i}\varepsilon_{0}+\stackrel[i=1]{n}{\Sigma}\Delta B_{i}\delta\varepsilon_{i}=\stackrel[i=1]{n}{\Sigma}\Delta B_{i}\delta\varepsilon_{i}$.
Therefore the contribution from the known modes is derived as

\begin{equation}
\frac{\Delta\sigma_{\mathrm{ZH}}}{\sigma_{\mathrm{ZH}}}=\frac{\Delta\overline{\varepsilon}}{\overline{\epsilon}}=\sqrt{\stackrel[i=1]{n}{\Sigma}\Delta B_{i}^{2}\left(\frac{\varepsilon_{i}}{\varepsilon_{0}}-1\right)^{2}}\,.\label{eq:BR4}
\end{equation}

Scenario C is the most realistic as we will certainly have branching
ratio measurements from both the LHC and the ILC itself for a wide
range of Higgs decay modes.

From the above formulation, it is apparent that the key to maintaining
model independence is to minimize the discrepancies in signal efficiency
between decay modes. This is exactly the guideline for designing the
data selection methods in Section \ref{sec:Event-Selection}, while
still allowing them to achieve high precision of $\sigma_{\mathrm{ZH}}$
and $M_{\mathrm{H}}$. To cover a large number of Higgs decay modes
and monitor their efficiencies, high statistics signal samples ($\sim$
40k events) are produced for each major SM decay mode ($\mathrm{H\rightarrow\mathrm{bb}}$,
cc, gg, $\tau\tau$, $\mathrm{WW^{*}}$, $\mathrm{ZZ^{*}}$, $\gamma\gamma$,
$\gamma\mathrm{Z}$), and for each beam polarisation and center-of-mass
energy, so that the relative statistical error of each efficiency
is below 0.2\% in the end for any channel. As for the analysis strategies,
from the very beginning, the isolated lepton selection mentioned in
Section \ref{sub:Isolated-Lepton-Finder} is tuned to take into account
the fact that each decay mode has a different density of particles
surrounding the leptons from Z boson decay. Then, as explained in
Section \ref{sub:pair}, the isolated leptons are carefully paired
to minimize the chance of including leptons from Higgs decay into
the pair\cite{Yan:2016trc}. Following these signal selection
processes, the cuts on $M_{\mathrm{l^{+}l^{-}}}$, $p_{\mathrm{T}}^{\mathrm{l^{+}l^{-}}}$,
BDT, and $M_{\mathrm{rec}}$ are designed to use only kinematical
information from the selected leptons so as to avoid introducing bias
to the efficiencies of individual Higgs decay modes. Even though the
$\cos\left(\theta_{\mathrm{missing}}\right)$ cut, which counts the
missing momentum from the whole event but, in principle uses information
of particles from Higgs decay, it is designed to be so loose that
there is almost no effect on signal efficiency, while 2-fermion backgrounds
can still be suppressed effectively. The $E_{\mathrm{vis}}$ cut will
not introduce additional bias either, as it simply categorizes the
events into visible or invisible Higgs decay, as mentioned in Section
\ref{sub:Background-Rejection}. More details on the data selection
strategies regarding model independence can be found in\cite{Yan:2016trc}.
Table \ref{tab:Mode Eff} shows the efficiencies of each decay mode
after each cut for the $\mathrm{\mu^{+}\mu^{-}H}$ channel
at $\sqrt{s}$=250 GeV.

\begin{table*}
\centering
\caption{The BR values and efficiencies of the major SM Higgs decay modes,
after each data selection step, shown here for the case of the $\mathrm{\mu^{+}\mu^{-}H}$
channel and $\mathrm{e_{L}^{-}e_{R}^{+}}$ at $\sqrt{s}$=250 GeV.
The statistical uncertainties on these values are below 0.14\%.}
\label{tab:Mode Eff}
\begin{tabular}{|c|c|c|c|c|c|c|c|c|}
\hline
$\mathrm{H\rightarrow XX}$ & bb & cc & gg & $\tau\tau$ & $\mathrm{WW^{*}}$ & $ZZ^{*}$ & $\gamma\gamma$ & $\gamma Z$\tabularnewline
\hline
BR (SM) & 57.8\% & 2.7\% & 8.6\% & 6.4\% & 21.6\% & 2.7\% & 0.23\% & 0.16\%\tabularnewline
\hline
\hline
Lepton Finder & 93.70\% & 93.69\% & 93.40\% & 94.02\% & 94.04\% & 94.36\% & 93.75\% & 94.08\%\tabularnewline
\hline
Lepton ID+Precut & 93.68\% & 93.66\% & 93.37\% & 93.93\% & 93.94\% & 93.71\% & 93.63\% & 93.22\%\tabularnewline
\hline
$M_{\mathrm{l^{+}l^{-}}}\in[73, 120]$ GeV & 89.94\% & 91.74\% & 91.40\% & 91.90\% & 91.82\% & 91.81\% & 91.73\% & 91.47\%\tabularnewline
\hline
$p_{\mathrm{T}}^{\mathrm{l^{+}l^{-}}}\in[10, 70]$ GeV & 89.94\% & 90.08\% & 89.68\% & 90.18\% & 90.04\% & 90.16\% & 89.99\% & 89.71\%\tabularnewline
\hline
$\left|\cos\theta_{\text{miss}}\right|< 0.98$ & 89.94\% & 90.08\% & 89.68\% & 90.16\% & 90.04\% & 90.16\% & 89.91\% & 89.41\%\tabularnewline
\hline
BDT > - 0.25 & 88.90\% & 89.04\% & 88.63\% & 89.12\% & 88.96\% & 89.11\% & 88.91\% & 88.28\%\tabularnewline
\hline
$M_{\mathrm{rec}}\in [110, 155]$ GeV & 88.25\% & 88.35\% & 87.98\% & 88.43\% & 88.33\% & 88.52\% & 88.21\% & 87.64\%\tabularnewline
\hline
\end{tabular}
\end{table*}

Table \ref{tab:Mode Eff} shows no discrepancy in efficiencies beyond
1\%, which demonstrates model independence at a level of better than
0.5\% based on the most conservative scenario B. The bias is even
smaller at higher center-of-mass energies\cite{Yan:2016trc}.
For example, at $\sqrt{s}$=500 GeV, no bias exists beyond the MC
statistical error (< 0.2\%) for any mode. Regarding the most realistic scenario
C, the estimation of potential bias is obtained as follows (using
Equations \ref{eq:BR3} and \ref{eq:BR4}). The known modes are assumed
to be $\mathrm{H\rightarrow\mathrm{bb}}$, cc, gg, $\tau\tau$, $\mathrm{WW^{*}}$,
$\mathrm{ZZ^{*}}$, $\gamma\gamma$, and $\gamma\mathrm{Z}$, since they
will be measured at the LHC or the ILC\cite{Dawson:2013bba,Asner:2013psa}.
The total branching ratio for the unknown modes ($B_{x}$) is assumed
to be 10\%, based on the estimation of the 95\% C.L. upper limit for
branching ratio of BSM decay modes from the HL-LHC\cite{Dawson:2013bba}.
In fact this assumption is rather conservative, because at the ILC
the upper limit for BSM decay will be greatly improved and in general
any decay mode with a few percent branching ratio shall be directly
measured. Since the characteristics of any exotic decay mode are expected
to fall within the wide range of known decay modes being directly
investigated, we obtain $\delta\varepsilon_{\mathrm{max}}$ by assuming
that the efficiencies of the unknown modes will lie in the range of
the efficiencies of known modes; this is, for example, -0.68\% from
the $\gamma\mathrm{Z}$ mode in the case of the channel shown in Table
\ref{tab:Mode Eff}. Then for the known modes, each $B_{i}$ is scaled
from their SM values by 90\%, following which $\varepsilon_{0}$ is
obtained straightforwardly from $B_{i}$ and $\epsilon_{i}$. Each
$\Delta B_{i}$ is taken conservatively from the largest uncertainties
predicted from the HL-LHC measurements\cite{Dawson:2013bba} with exceptions
of the $\mathrm{H\rightarrow cc}$ and gg modes which are very difficult
to obtain at the HL-LHC and thus are obtained from the predictions
for the ILC\cite{Asner:2013psa}.

Table \ref{tab:ModeBias} shows for all center-of-mass energies and
polarizations in this analysis the relative bias on $\sigma_{\mathrm{ZH}}$,
which is below 0.1\% for the $\mathrm{\mu^{+}\mu^{-}H}$
channel and 0.2\% for the $\mathrm{\mathrm{e}^{+}\mathrm{e}^{-}X}$
channel. The maximum contribution to the residual bias comes from
either the $\mathrm{H\rightarrow\gamma\gamma}$ mode or the $\mathrm{H\rightarrow\gamma Z}$
mode.

From the the above and results in Table \ref{tab:ModeBias}, we conclude
that the model independence of $\sigma_{\mathrm{ZH}}$ measurement
at the ILC using Higgsstrahlung events $\mathrm{e^{+}e^{-}\rightarrow ZH\rightarrow l^{+}l^{-}H}$
($\mathrm{l}$ = e or $\mu$) is demonstrated to a level well below
even the smallest statistical $\sigma_{\mathrm{ZH}}$ uncertainties
expected from the leptonic recoil measurements in the full H20 run,
by a factor of 5.

\begin{table}
\centering
\caption{The relative bias on $\sigma_{\mathrm{ZH}}$ evaluated for each center-of-mass
energy and polarization.}
\label{tab:ModeBias}
\begin{tabular}{|c|c|c|c|c|c|c|}
\hline
$\sqrt{s}$ & \multicolumn{2}{c|}{250 GeV}  & \multicolumn{2}{c|}{350 GeV}  & \multicolumn{2}{c|}{500 GeV}  \tabularnewline
\hline
$\mathrm{l^{+}l^{-}}\mathrm{H}$ & $\mathrm{\mu^{+}\mu^{-}X}$  & $\mathrm{e}^{+}\mathrm{e}^{-}X$ & $\mathrm{\mu^{+}\mu^{-}X}$  & $\mathrm{e}^{+}\mathrm{e}^{-}X$ & $\mathrm{\mu^{+}\mu^{-}X}$  & $\mathrm{e}^{+}\mathrm{e}^{-}X$\tabularnewline
\hline
\hline
$\mathrm{e_{L}^{-}e_{R}^{+}}$ & 0.08\% & 0.19\% & 0.04\% & 0.11\% & 0.05\% & 0.09\%\tabularnewline
\hline
$\mathrm{e_{R}^{-}e_{L}^{+}}$ & 0.06\% & 0.13\% & 0.00\% & 0.12\% & 0.02\% & 0.02\%\tabularnewline
\hline
\end{tabular}
\end{table}

\section{Summary and Conclusions}
\label{sec:Summary-and-Conclusion}

Precise measurements of the absolute cross section ($\sigma_{\mathrm{ZH}}$)
in a model independent way and the Higgs boson mass ($M_{\mathrm{H}}$)
at the ILC are essential for providing sensitivity to new physics
beyond the Standard Model. By applying the recoil technique to the
Higgsstrahlung process with the Z boson decaying to a pair of electrons
or muons, the precision of the measurement of $\sigma_{\mathrm{ZH}}$
and $M_{\mathrm{H}}$ has been evaluated for the proposed ILC run
scenario based on the full simulation of the ILD. A clear comparison
has been established between three center of mass energies $\sqrt{s}$
= 250, 350, and 500 GeV, and two beam polarizations $\left(P\mathrm{e^{-}},P\mathrm{e^{+}}\right)$
=($-$80\%, +30\%) and (+80\%, $-$30\%). The results presented contribute
to further optimization of the ILC run scenario.

Assuming an integrated luminosity of 250 $\mathrm{fb^{-1}}$ at $\sqrt{s}$
= 250 GeV, where the best lepton track momentum resolution is obtainable,
$\sigma_{\mathrm{ZH}}$ and $M_{\mathrm{H}}$ can be determined with
a precision of 2.5\% and 37 MeV for $\mathrm{e_{L}^{-}e_{R}^{+}}$
and 2.9\% and 41 MeV for $\mathrm{e_{R}^{-}e_{L}^{+}}$, respectively.
Regarding a 20 year ILC physics program, the expected precisions for
the $\mathrm{HZZ}$ coupling and $M_{\mathrm{H}}$ are 0.4\% and 14
MeV, respectively. Precision can be further improved by combining
with hadronic recoil results.

Methods of signal selection and background rejection are optimized
to not only achieve the high precision of $\sigma_{\mathrm{ZH}}$
and $M_{\mathrm{H}}$, but also to minimize the bias on the measured
$\sigma_{\mathrm{ZH}}$ due to discrepancy in signal efficiencies
among Higgs decay modes. As a result, the model independence of the
leptonic recoil measurement has been demonstrated for the first time
to the sub-percent level; the relative bias on $\sigma_{\mathrm{ZH}}$
is below 0.1\% for the $\mathrm{\mu^{+}\mu^{-}H}$ channel
and 0.2\% for the $\mathrm{e^{+}e^{-}X}$ channel,
which is at least five times smaller than even the smallest $\sigma_{\mathrm{ZH}}$
statistical uncertainties expected from the leptonic recoil measurements
in a full 20 years ILC physics program.

\begin{acknowledgments}
The authors would like to thank T. Barklow and colleagues from the
ILD Concept Group for their help in realizing this paper; in particular,
A. Miyamoto, C. Calancha, and M. Berggren for their work in generating
the Monte-Carlo samples. J. Yan would also like to thank S. Komamiya
for helpful discussions and suggestions. This work has been partially
supported by JSPS Grants-inAid for Science Research No. 22244031 and
the JSPS Specially Promoted Research No. 23000002.
\end{acknowledgments}

\nocite{apsrev41Control}
\bibliographystyle{apsrev4-1}
\bibliography{recoil}
\end{document}